\documentclass[journal]{IEEEtran}
\IEEEoverridecommandlockouts
\usepackage{cite}
\usepackage{amsmath,amssymb,amsfonts,amsthm}
\usepackage{graphicx}
\usepackage{textcomp}
\usepackage[table]{xcolor}
\usepackage{booktabs}
\usepackage{array}
\usepackage{multirow}
\usepackage[hidelinks]{hyperref}

\definecolor{hl}{RGB}{200,235,200}
\definecolor{inp}{RGB}{205,225,248}
\definecolor{dvar}{RGB}{253,220,180}
\definecolor{gobj}{RGB}{226,236,248}
\definecolor{grob}{RGB}{248,228,228}
\definecolor{gnet}{RGB}{228,242,228}
\definecolor{gbud}{RGB}{246,240,222}
\usepackage{url}
\usepackage{enumitem}
\usepackage{algorithm}
\usepackage[noend]{algpseudocode}
\usepackage{tikz}
\usetikzlibrary{arrows.meta,positioning,calc}
\usepackage{pgfplots}
\usepgfplotslibrary{groupplots}
\pgfplotsset{compat=1.17}
\newtheorem{theorem}{Theorem}
\newtheorem{lemma}{Lemma}

\newtheorem{corollary}{Corollary}
\newtheorem{definition}{Definition}
\newtheorem{remark}{Remark}
\newcommand{\F}{\mathbb{F}_q}
\newcommand{\Np}{N^{+}}
\newcommand{\Ncp}{N^{+}_c}
\newcommand{\Nbar}{\bar N_c}
\newcommand{\Nbarp}{\bar N^{+}_c}

\newcommand{\Ncell}{N_{r,c}}
\newcommand{\hQ}{h_Q}

\begin{document}
\bstctlcite{IEEEexample:BSTcontrol}

\title{TCDA: Tunable Coded Distributed Arrays \\for Data Availability Sampling}

\author{Duc A. Tran, Dung Truong, and Duy Le%
\thanks{The authors are with the Department of Computer Science, University of Massachusetts, Boston, MA, USA. Email: duc.tran@umb.edu}}

\maketitle

\begin{abstract}
Data availability sampling (DAS) lets blockchain nodes verify a block's data availability by downloading only a few random pieces of it. Its network layer must disseminate each block in time for validators to attest (PUT), store it, and serve any piece on request (GET). A trustless DAS must sustain any number of Byzantine peers. The only known design with this guarantee, Robust Distributed Arrays (RDA), achieves it by full replication: each group of nodes responsible for part of a block stores that part in full, so a single good node per group suffices. The price to pay is a PUT load on the critical path to finality that grows with the network and cannot be tuned, hence the data may not reach validators before the attestation deadline. Coding is the natural remedy, but it was not known when a coded layer remains robust against an unbounded adversary, what it costs, or what it gains.
We propose TCDA, a coded layer that is provably robust given enough good nodes, wherever they lie, with a tunable knob: how much of its assigned data each node keeps. Turning the knob trades PUT and storage costs against GET cost. It can be set per data type and adjusted as the good-node population changes. 
At Ethereum's parameters, our evaluation for a $32$\,MiB block at the recommended uplink shows that RDA misses the attestation deadline at any network size, whereas TCDA can be tuned to meet it from $868$ good nodes up to the $40{,}000$ tested, storing $1.7$ to $32$ times less per node than replication, at the price of $2.5$ to $7.7$ round-trips per sample instead of about one.

\end{abstract}


\section{Introduction}
\label{sec:intro}

Rollup-centric blockchains publish large blobs whose only purpose is to be \emph{available} for a bounded window. Data availability sampling (DAS)~\cite{fraud-proofs,hall2023foundations} lets a node convince itself that a block is available by fetching a few random symbols of its erasure-coded extension, without downloading the block. Ethereum introduced blobs with EIP-4844~\cite{eip4844} and has sampled them with PeerDAS~\cite{peerdas,peerdas-doc} since its Fusaka upgrade~\cite{fusaka}. The cryptographic layer of DAS (erasure codes, commitments, samplers) is well understood~\cite{hall2023foundations,hall2024frida,zoda}. The \emph{networking} layer is harder because it has to do three things while an unknown and possibly large set of peers is Byzantine. First, the PUT protocol must spread every block's data across the network before validators attest; its bandwidth and latency sit on the critical path to finality. Second, storage must keep each node's share for the whole retention window; this burden falls on every node for every block and decides who can afford to participate. Third, the GET protocol must let a sampling query reach the requested symbol in a few round-trips. A tradeoff is inevitable: the redundancy that makes GET trivial is exactly what makes PUT and storage expensive.

\textbf{Robust Distributed Arrays (RDA).} RDA~\cite{rda} is the first provably secure networking layer for DAS, recently proposed by Ethereum Foundation. Nodes are hashed into a $K_1\times K_2$ grid and form a clique with every peer in their row and in their column. The extended block is cut into $K_2$ chunks, each  fully replicated at an assigned column. Availability is guaranteed provided that \emph{every column contains at least one good node}. This assumption concerns the good nodes only, so RDA remains robust under any number of Byzantine ones.

RDA offers trustless DAS, but it can fail deadline and storage constraints. Consider Ethereum scale: a $12$\,s slot in which validators attest about $4$\,s in~\cite{paradigm-slot} and an $18$-day retention window ($131{,}072$ blocks)~\cite{eip4844}, and a $32$\,MiB extended object ($8$\,MiB of blob data) on a $15\times32$ grid.  First, the PUT misses the attestation deadline. On the smallest network satisfying RDA robustness assumption, $965$ good nodes and as many Byzantine ones, each relay must forward about $60$\,MiB per object, i.e., $127$\,Mbit/s over $4$\,s, five times the $25$\,Mbit/s recommended for an attesting node~\cite{eip7870}. 
 Second, storage is expensive: each node needs $128$\,GiB at any time, twice the amount PeerDAS \cite{peerdas}, the current DAS deployment on Ethereum, stores per node.

\textbf{Motivation.} The example reflects a pain point of the whole field, not only of RDA. The PUT sits on the critical path to finality, yet no existing DAS design offers both a dissemination load that a deployment can size against its slot \emph{and} provable robustness against any number of Byzantine peers. Today's designs that are provably robust fix the load at its maximum and that bound the load give up the proof.

Coding is the natural remedy. If each column node keeps coded pieces worth $1/\kappa$ of its chunk, the column dissemination and the per-node storage shrink $\kappa$-fold, as shown in our earlier work~\cite{cda}, which applies random linear network coding (RLNC)~\cite{rlnc}. Coding, however, raises issues that replication never had:
1) \emph{The availability guarantee breaks:} a good node in every column is no longer enough, since such a node does not hold the whole chunk, and so retrieval is not guaranteed;
2) \emph{A network may not meet the new robustness condition:} a new condition is needed and a deployment must be able to tell from its design inputs whether it is met;
3) \emph{Byzantine nodes gain a new lever:} they are unbounded in number and can supply valid but useless pieces, which may degrade what the good nodes keep; and
4) \emph{The savings are paid for elsewhere:} as no single good node can answer a sampling query, each GET must combine pieces from several good nodes and costs more.

\textbf{Problem.} We address all of these issues and make coding provably robust against an unbounded adversary. We ask: \emph{without any good-node assumption, exactly when is a coded array robust, what does it cost, and what does it offer that replication cannot?} We break this into three questions.
\begin{enumerate}[leftmargin=*,label=\textbf{Q\arabic*}]
\item \emph{Under exactly what condition is a coded array robust?} The condition must be necessary and sufficient, hold under arbitrarily many bad nodes, and be stated only as the number of good nodes the network needs, wherever they lie. Unlike RDA's per-column assumption, which no participant can observe, this number is a condition computed from the robustness requirement and the code.
\item \emph{What does coding trade, and is it reasonable for blockchain DAS?} The PUT must meet the attestation deadline in every slot and storage burdens every node, while the GET should be as fast as possible. We need to state this trade-off exactly, so that the code can be tuned to different DAS requirements.
\item \emph{Does coding offer anything beyond lower costs?} Lower costs are what coding is expected to deliver. Yet storing coded pieces instead of full copies also changes what each node contributes to the system. We ask whether this brings benefits that replication cannot provide by design, while never doing worse.
\end{enumerate}

\textbf{Contributions.} We answer these questions with TCDA (Tunable Coded Distributed Array). TCDA keeps RDA's grid but not its per-column assumption, and changes the PUT and GET protocols. Instead of a full chunk, a column node stores only $Q \le \kappa$ coded pieces. This $Q$, the \emph{retention depth}, is the single knob tuning the PUT and storage costs against the GET cost. Fig.~\ref{fig:vs} contrasts TCDA with RDA. Our key contributions:

\begin{figure}[t]
\centering
\begin{tikzpicture}[x=1cm,y=0.88cm,every node/.style={font=\scriptsize}]
\def\slot{0.28}
\node[font=\small\bfseries] at (1.55,3.55) {RDA};
\foreach \k/\g in {0/1,1/0,2/1,3/0,4/1}{
  \pgfmathsetmacro{\y}{2.9-0.55*\k}
  \ifnum\g=1 \node[circle,draw,fill=green!50,inner sep=1.4pt] at (0.25,\y) {}; \else \node[circle,draw,fill=red!45,inner sep=1.4pt] at (0.25,\y) {}; \fi
  \foreach \p in {0,...,3}{\fill[orange!55] (0.55+\p*\slot*1.4,\y-0.13) rectangle ++(\slot*1.4,0.26);}
  \draw (0.55,\y-0.13) rectangle ++(4*\slot*1.4,0.26);
  \ifnum\g=0 \draw[red!70!black,thick] (0.5,\y-0.17) -- (2.17,\y+0.17); \fi
}
\node[align=center] at (1.35,0.12) {every node: \textbf{whole} chunk\\ ($\kappa$ of $\kappa$ pieces)};
\node[align=center] at (1.35,-0.62) {
GET: one good contact};
\node[font=\small\bfseries] at (5.35,3.55) {TCDA ($\kappa{=}4$, $Q{=}2$)};
\foreach \k/\g/\ca/\cb in {0/1/blue!55/teal!55,1/0/violet!45/cyan!45,2/1/magenta!45/olive!50,3/0/brown!45/lime!55,4/1/purple!45/yellow!70!orange}{
  \pgfmathsetmacro{\y}{2.9-0.55*\k}
  \ifnum\g=1 \node[circle,draw,fill=green!50,inner sep=1.4pt] at (4.05,\y) {}; \else \node[circle,draw,fill=red!45,inner sep=1.4pt] at (4.05,\y) {}; \fi
  \fill[\ca] (4.35,\y-0.13) rectangle ++(\slot*1.4,0.26);
  \fill[\cb] (4.35+\slot*1.4,\y-0.13) rectangle ++(\slot*1.4,0.26);
  \draw (4.35,\y-0.13) rectangle ++(2*\slot*1.4,0.26);
  \draw[gray!60,densely dotted] (4.35+2*\slot*1.4,\y-0.13) rectangle ++(2*\slot*1.4,0.26);
  \ifnum\g=0 \draw[red!70!black,thick] (4.3,\y-0.17) -- (5.2,\y+0.17); \fi
}
\draw[-{Latex[length=1.3mm]},green!45!black,thick] (5.35,1.8) to[out=0,in=0,looseness=1.3] (5.35,2.85);
\node[align=left,anchor=west,font=\tiny] at (6.05,2.35) {2 good nodes\\$\to$ rank $4=\kappa$};
\node[align=center] at (5.35,0.12) {every node: $Q$ \textbf{coded} pieces\\
($Q/\kappa$ of the chunk)};
\node[align=center] at (5.35,-0.62) {
GET: $\lceil\kappa/Q\rceil{-}1$ more good contacts};
\draw[gray!50] (3.2,-1.0) -- (3.2,3.75);
\end{tikzpicture}
\caption{RDA vs.\ TCDA on one column of five nodes, three good. RDA stores the whole chunk at every node, so one good node suffices for a GET. TCDA stores $Q$ of $\kappa$ coded pieces per node (colors: distinct coefficients), so a GET needs to contact multiple good nodes.}
\label{fig:vs}
\end{figure}
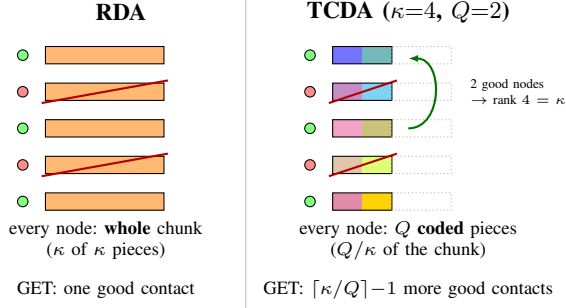

\begin{itemize}[leftmargin=*]
\item \emph{Novelty}: TCDA is the first provably robust coded DAS layer. Its robustness is conditioned only on  the number $\Np$ of good nodes in the network, whereas RDA's assumption is made for every column which is harder to enforce. The design is not trivial: with naive coding, one Byzantine relay can make every good node keep the same useless pieces, which is not to mention deadline and storage constraints.
\item \emph{Robustness} (Q1): We derive the exact probability that sampling a symbol fails and, from it, the necessary and sufficient condition for a given $\delta$-robustness: $\Np\ge\Np_\delta(Q)$, a single numbercomputed from $\delta$ and the code. We make no assumption where they lie; a column can have no good node. 
\item \emph{Tunability} (Q2): PUT and storage fall to exactly $Q/\kappa$ of RDA's. The price is paid at GET, which takes a second hop into the column: about $\log_2\big((\lceil\kappa/Q\rceil-1)/\eta\big)+1$ extra round-trips, where $\eta$ is the fraction of good nodes. Bad nodes thus slow a GET only logarithmically and add almost nothing to its traffic. A deployment can tune $Q$ to its deadline and sampling needs, per data type if needed, and is never worse off than with RDA, which is TCDA at $Q=\kappa$.
\item \emph{Optimal design} (Q2): With the attestation deadline and a storage budget, we choose retention depth $Q$ jointly with the grid $(K_1,K_2)$ for the fastest GET. On every grid the optimal $Q$ has a closed form, the largest one within the budgets, and so the joint design reduces to a finite scan over grids, which is fast to find a solution. We also give the range of the number of good nodes  and the largest object size for which a feasible design exists.
\item \emph{Scalability} (Q3): Coding changes how the network responds to growth. In RDA, as every new node adds traffic to PUT load, a growing network will eventually miss the deadline. In TCDA, more good nodes allow a smaller retention depth $Q$, which absorbs the growth: a grid carries $\kappa/Q$ times more nodes within the same PUT budget. Each bad node also adds only $Q/\kappa$ as much PUT load as in RDA, and robustness holds however many of them join.
\item \emph{Validation}: We evaluate TCDA at Ethereum's timing and bandwidth with both static and dynamic networks, against the two most relevant baselines: RDA and PeerDAS~\cite{peerdas}, the current DAS deployed on Ethereum. At the recommended $25$\,Mbit/s uplink, for object size of $32$\,MiB, RDA misses the attestation deadline at every network size, while TCDA can be tuned to meet it from $868$ good nodes up to the $40{,}000$ we tested,  storing $1.7$ to $32$ times less per node than replication, at the price of $2.5$ to $7.7$ round-trips per sample instead of about one. TCDA stays robust under any number of Sybils, where PeerDAS, in our model, does not. TCDA maintains availability under steady churn, and even after a sudden outage of up to $55\%$ of the good nodes. 
\end{itemize}

Section~\ref{sec:related} reviews related work. Sections~\ref{sec:model} and~\ref{sec:protocol} give the system model and the TCDA protocol. Sections~\ref{sec:security} to \ref{sec:design} analyze its robustness, formulate costs, and  derive an optimal design. Section~\ref{sec:eval} reports evaluation results. Section \ref{sec:discussion} discusses limitations and open problems of TCDA. Section \ref{sec:conclusion} concludes the paper.

\section{Related Work}
\label{sec:related}

\textbf{DAS and its cryptographic layer.} Al-Bassam et al.~\cite{fraud-proofs} introduced availability checks by random sampling of a 2D Reed--Solomon extension, later deployed by Celestia~\cite{lazyledger}. Yu et al.~\cite{cmt} use LDPC codes in a coded Merkle tree to shorten the proofs of incorrect coding. Hall-Andersen, Simkin, and Wagner defined DAS as a cryptographic primitive and analyzed tensor-code constructions~\cite{hall2023foundations,hall2024frida}; ZODA~\cite{zoda} and~\cite{accidental} reduce the commitment overhead. These works decide \emph{what} is committed and sampled, and treat the network as a black box that returns the symbol at a queried position~\cite{rda}. We address that black box: \emph{where} the symbols are stored and how they are fetched.

\textbf{Networking for DAS.} Kr\'ol et al.~\cite{das-p2p} set out what DAS asks of a P2P network and why classical DHTs such as Kademlia~\cite{kademlia} and Chord~\cite{chord} do not deliver it as they are. A lookup takes several hops, storing the samples in a Kademlia DHT is too heavy for the slot~\cite{dht-das}, and the robust DHT variants need most nodes to be good~\cite{rda}. PeerDAS~\cite{peerdas,peerdas-doc}, deployed on Ethereum with Fusaka update~\cite{fusaka}, extends each blob in one dimension, spreads the resulting columns over GossipSub~\cite{gossipsub} topics, and has each node keep a few of them. PANDAS~\cite{pandas} targets Ethereum's slot timing with builder-driven dissemination to deterministically assigned nodes. Honeybee~\cite{honeybee} samples peers by verifiable random walks and tolerates a Byzantine majority, but it is a peer-sampling primitive, not a storage layer. These designs bound what a node forwards, yet none proves that sampling succeeds against an arbitrary number of Byzantine peers. In an open network identities are free~\cite{sybil}, and Section~\ref{sec:eval} shows what this does to PeerDAS in our model.

RDA~\cite{rda} is the only design with such a proof. It is the latest development by the Ethereum Foundation as a response to the limitations of multi-hop dissemination (gossip and DHT) in adversarial settings. 
It, organized as a grid network, is fast for sampling and robust under any number of Byzantine peers provided that every column holds a good node. As pointed out in the previous section, RDA cannot scale as more nodes join and easily violate deadline constraints in today's blockchains.

\textbf{Coded storage and dispersal.} RLNC~\cite{rlnc} recovers data from any $\kappa$ independent coded pieces, and regenerating codes~\cite{dimakis} apply network coding to storage, trading storage against repair bandwidth; neither considers Byzantine nodes. Verifiable information dispersal~\cite{avid} stores an erasure-coded file on $n$ servers, robustly but only up to an $1/3$ Byzantine rate; in contrast, our Byzantine model assumes only a minimum number of honest nodes, allowing for any number of Byzantine ones. With Byzantine relays, coding also invites pollution, i.e., forged coded pieces. Homomorphic signatures~\cite{nc-sig} let a node verify a coded piece without decoding, and in DAS the KZG commitments~\cite{kzg} already serve this purpose.  CDA~\cite{cda}, our earlier work, provides evidence that RLNC when applied to RDA's grid with such a KZG-homomorphic verification layer can cut costs in PUT and storage. To the best of our knowledge, no prior work on coded DAS, including CDA, proves robustness to the unbounded Byzantine model. Our research is the first effort to achieve this, and, especially, we offer a provably robust coded DAS design that  can be tuned for different deadline and storage constraints.

\section{System Model}
\label{sec:model}

The system model has four parts: the data and its coding, the network and its adversary, the robustness requirement, and the condition on the good nodes.

\subsection{Data and coding}
\label{sec:prelim}
\textbf{Object and tensor code.} An \emph{object}, e.g., a block, is a $D\times D$ matrix whose symbols each consist of $\kappa$ elements of $\F$, the scalar field of BLS12-381 ($q\approx2^{255}$). Equivalently, it has $\kappa$ \emph{layers}, each a $D\times D$ matrix over $\F$. Each layer is extended by a 2D Reed--Solomon (tensor) code to an $M\times M$ codeword, typically with $M=2D$~\cite{hall2023foundations}, giving the extended object $\hat{\mathcal B}=(X_1,\dots,X_\kappa)$. Its $M^2$ \emph{positions} $(r_1,c_1)\in[M]^2$ each hold one symbol, and they are what clients GET and sample. Block $\hat{\mathcal B}$ can be fully reconstructed provided that fewer than $(M-D+1)^2$ symbols are missing; for $M=2D$ the code tolerates a missing fraction just above $1/4$.

\textbf{Commitments.} Data column $j$ of layer $i$ is the evaluation vector of a polynomial of degree less than $D$, committed with KZG~\cite{kzg} as $\mathrm{com}_{i,j}$, and $\mathrm{com}$ denotes the hash of all $\kappa M$ commitments. KZG is additively homomorphic: $\mathsf{Com}(\sum_i g_if_i)=\sum_i g_i\,\mathsf{Com}(f_i)$, and an opening proof of $\sum_i g_if_i$ at a point is $\sum_i g_i$ times the individual proofs. Binding holds under $q$-SDH~\cite{bb-signatures}. A node that holds all $M$ evaluations of a committed polynomial can compute an opening at any point itself (amortized in $O(M\log M)$ by FK20~\cite{fk20}).


\subsection{Network and adversary}
\label{sec:net}

\begin{figure}[t]
\centering
\begin{tikzpicture}[x=0.95cm,y=0.8cm,every node/.style={font=\scriptsize},
  good/.style={circle,draw,fill=green!50,inner sep=1.4pt},
  bad/.style={circle,draw,fill=red!45,inner sep=1.4pt},
  boot/.style={rectangle,draw,fill=yellow!60,inner sep=2pt}]
\foreach \i in {0,...,3}{\foreach \j in {0,...,5}{\draw[gray!55] (\j,\i) rectangle ++(1,1);}}
\foreach \j in {1,...,6}{\node[gray] at (\j-0.5,4.25) {\tiny $c{=}\j$};}
\node[gray,rotate=90] at (-0.95,2) {\tiny rows $1..K_1$};
\fill[blue!10] (0,2) rectangle (6,3);
\fill[orange!18] (2,0) rectangle (3,4);
\fill[orange!40] (2,2) rectangle (3,3);
\foreach \x/\y/\s in {0.3/3.6/good,0.7/3.3/bad,1.5/3.5/bad,2.4/3.7/good,3.3/3.4/good,3.7/3.7/bad,4.5/3.5/good,5.3/3.6/bad,5.7/3.3/good,
  0.5/2.6/bad,1.3/2.4/bad,1.7/2.7/bad,3.5/2.5/bad,4.3/2.6/good,4.7/2.3/bad,5.5/2.5/good,
  0.4/1.5/bad,1.5/1.6/bad,2.3/1.3/bad,2.7/1.7/good,3.5/1.5/good,4.5/1.4/bad,5.4/1.6/bad,
  0.6/0.5/bad,1.4/0.4/bad,2.5/0.5/bad,3.3/0.6/bad,3.7/0.3/good,4.5/0.6/good,5.5/0.4/good}
  {\node[\s] at (\x,\y) {};}
\node[good,inner sep=1.8pt,line width=0.9pt] (u) at (2.5,2.5) {};
\node[font=\tiny] at (2.5,2.18) {$u$};
\foreach \x/\y in {0.5/2.6,1.3/2.4,1.7/2.7,3.5/2.5,4.3/2.6,4.7/2.3,5.5/2.5}{\draw[blue!60,thin] (u) -- (\x,\y);}
\foreach \x/\y in {2.4/3.7,2.3/1.3,2.7/1.7,2.5/0.5}{\draw[orange!70!black,thin] (u) -- (\x,\y);}
\foreach \i in {0,...,3}{\node[boot] (b\i) at (6.55,\i+0.5) {};\draw[gray,densely dotted] (6,\i+0.5) -- (b\i);}
\node[font=\tiny,align=left,anchor=west] at (6.72,2) {bootstrap\\(one per row;\\joins every\\row subnet)};
\foreach \j/\g in {1/1,2/0,3/3,4/3,5/3,6/3}{\node at (\j-0.5,-0.3) {\tiny $\Ncp{=}\g$};}
\node[anchor=north west] at (-0.3,-1.05) {\tikz\node[good]{}; good\ \ \tikz\node[bad]{}; bad\ \ \tikz\node[boot]{}; bootstrap\ \ \textcolor{blue!60}{---} row links of $u$\ \ \textcolor{orange!70!black}{---} column links of $u$};
\end{tikzpicture}
\caption{The grid network: Each node $u$ keeps a connection to every peer in its row (blue) and its column (orange). Each row requires a bootstrap nodes. A column may contain no good node; e.g., column $c=2$. (Section~\ref{sec:assume}).}
\label{fig:grid}
\end{figure}
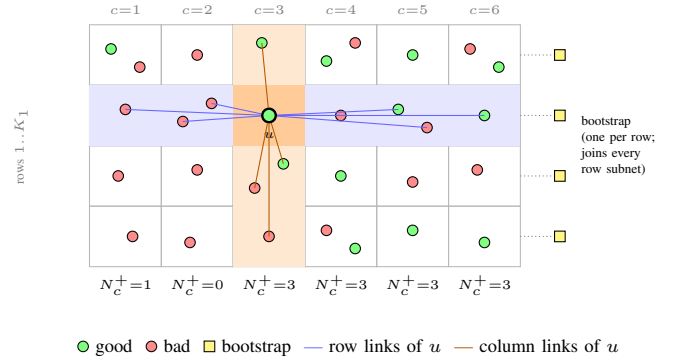

Like RDA~\cite{rda}, we structure the underlying P2P network as a $K_1 \times K_2$ grid (Fig.~\ref{fig:grid}). A random oracle maps each node $u$ to a cell $\mathrm{Cell}(u)=(r,c)\in[K_1]\times[K_2]$.  Each node keeps a connection to every node in its row or  column. For setting up new nodes, public bootstrap nodes are required such that every row must have one. A node is \emph{good} if it is honest and online for the relevant window; otherwise it is \emph{bad}. Bad nodes may deviate arbitrarily and collude. They see all messages and their messages can arrive before honest ones.

The network has $N$ nodes, $\Np$ of them good and $N^-=N-\Np$ bad. Column $c$ has $N_c$ nodes, $\Ncp$ of them good, and cell $(r,c)$ has $\Ncell$ nodes. For a node in row $r$, cell $(r,c)$ is its \emph{contact cell} for column $c$: its only one-hop route into that column. Node placement is uniform, so the mean column population is $\Nbar=N/K_2$ and the mean good count is $\Nbarp=\Np/K_2$. Let $\eta=\Np/N$ denote the good fraction. No bound is placed on $N^-$. Security of TCDA depends on $\Np$ alone.

\subsection{Requirement: $\delta$-robust sampling}
\label{sec:goal}
The network layer of DAS must respond to a random-position sampling of a published object with a correct symbol under a failure probability that still enables full reconstruction of the object. We state this requirement as follows.

\begin{definition}[$\delta$-robustness]
\label{def:robust}
A protocol is \emph{$\delta$-robust} if  a GET for any position never returns a wrong symbol and fails with probability at most $\delta$, whatever the bad nodes do.
\end{definition}

Each random sample must succeed with probability at least $1-\delta$ for the sampling argument of DAS~\cite{hall2023foundations}. The threshold $\delta$ is fixed and given by the code: up to $\delta M^2<(M-D+1)^2$  are allowed missing, so $\hat{\mathcal B}$ can be determined from the successful samples (Section~\ref{sec:prelim}). For the usual $M=2D$, any 75\% of positions suffice, and we take $\delta=1/4$. 

A GET can fail for three reasons: the data may be  {lost}, or   {unreachable}, or   {incomplete}. Unlike full replication where a single good node guards against loss and a single good responder completes a query, a coded design faces a harder task to achieve the same $\delta$-robustness. A node holds only a fraction of its part, so several good nodes must {jointly} hold enough. An answer must combine pieces from them, and as such a query can stall on bad responders. The adversary can also supply valid but useless pieces.

\subsection{Condition on the good nodes}
\label{sec:assume}
Two kinds of adversarial assumption are common: 1)
Byzantine protocols typically assume a threshold on the fraction of good nodes in the network, e.g., $\eta = 2/3$. No one can enforce this in an open network, since an adversary can add Sybil nodes to lower $\eta$ arbitrarily,   however many good nodes there are; and 
2) RDA avoids fractions but assumes, in advance of operation, that every column must hold at least one good node~\cite{rda}. This is a structural condition on the network grid: no participant can observe it, and a single column without a good node voids or weakens the guarantee. 

TCDA has a softer rule. It works upon a minimal number, $\Np_{\min}$, of good nodes in the network, not their fraction and not how they spread over the columns. Therefore, a column may contain no good node. The threshold  $\Np_{\min}$ is a  {bootstrap} condition computed at the design phase: given $\delta$-robustness   and the design parameters (grid sizing $K_1 \times K_2$ and retention depth $Q$), all fixed before deployment, we derive   $\Np_{\min} =\Np_\delta(Q)$ (Section~\ref{sec:security}). A deployment can go live once it has this many good nodes.

\section{The TCDA Protocol}
\label{sec:protocol}

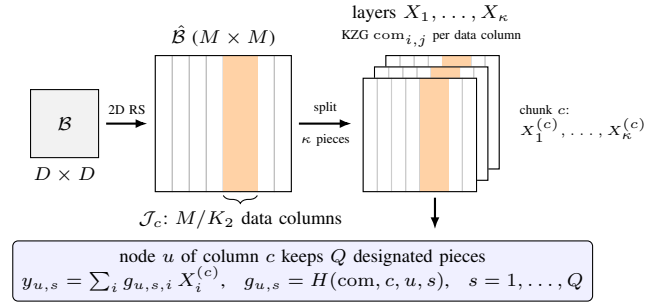
\begin{figure}[t]
\centering
\begin{tikzpicture}[x=1cm,y=1cm,every node/.style={font=\scriptsize}]
\draw[fill=gray!10] (0,0.45) rectangle (0.9,1.35);
\node at (0.45,0.9) {$\mathcal B$};
\node[below] at (0.45,0.45) {$D\times D$};
\draw[-{Latex[length=1.6mm]},thick] (1.0,0.9) -- node[above]{\tiny 2D RS} (1.55,0.9);
\foreach \j in {0,...,7}{\draw[gray!50] (1.65+0.225*\j,0) rectangle ++(0.225,1.8);}
\fill[orange!35] (1.65+0.225*4,0) rectangle ++(0.45,1.8);
\draw (1.65,0) rectangle (3.45,1.8);
\node[above] at (2.55,1.8) {$\hat{\mathcal B}$ ($M\times M$)};
\draw[decorate,decoration={brace,amplitude=3pt,mirror}] (1.65+0.9,-0.05) -- node[below=2pt]{$\mathcal{J}_c$: $M/K_2$ data columns} (1.65+1.35,-0.05);
\draw[-{Latex[length=1.6mm]},thick] (3.55,0.9) -- node[above]{\tiny split} node[below]{\tiny $\kappa$ pieces} (4.25,0.9);
\foreach \o in {0.3,0.15,0}{
  \begin{scope}[shift={(4.4+\o,\o)}]
  \fill[white] (0,0) rectangle (1.5,1.5);
  \foreach \j in {0,...,7}{\draw[gray!40] (0.1875*\j,0) rectangle ++(0.1875,1.5);}
  \fill[orange!35] (0.75,0) rectangle ++(0.375,1.5);
  \draw (0,0) rectangle (1.5,1.5);
  \end{scope}}
\node[above,align=center] at (5.3,1.8) {layers $X_1,\dots,X_\kappa$\\{\tiny KZG $\mathrm{com}_{i,j}$ per data column}};
\node[anchor=west,align=left] at (6.35,0.9) {\tiny chunk $c$:\\\tiny $X^{(c)}_1,\dots,X^{(c)}_\kappa$};
\draw[-{Latex[length=1.6mm]},thick] (5.35,-0.08) -- (5.35,-0.5);
\node[draw,rounded corners=2pt,fill=blue!6,align=center] at (3.6,-1.05) {node $u$ of column $c$ keeps $Q$ designated pieces\\ $y_{u,s}=\sum_i g_{u,s,i}\,X_i^{(c)}$,\quad $g_{u,s}=H(\mathrm{com},c,u,s)$,\quad $s=1,\dots,Q$};
\end{tikzpicture}
\caption{Data preparation: The 2D Reed--Solomon extension $\hat{\mathcal B}$ is split into $\kappa$ layers, one per symbol piece, and each data column of each layer is committed with KZG. Chunk $c$ is the block of $M/K_2$ contiguous data columns $\mathcal{J}_c$ (shaded) in all layers. It is stored by column $c$, where node $u$ keeps the $Q$ designated coded pieces $y_{u,1},\dots,y_{u,Q}$.}
\label{fig:enc}
\end{figure}

We present TCDA in five components:  1) preparation of a data object for publication, 2) how to publish it (PUT), 3) how to sample a symbol (GET), 4) how a node joins the network (JOIN), and 5) how to tune the parameters upon changing requirements (TUNE).

\subsection{Data Preparation}
The input to TCDA is the extended object $\hat{\mathcal B}=(X_1,\dots,X_\kappa)$. This is the 2D erasure-coded matrix of the original object $\mathcal B$ and its commitments $\mathrm{com}_{i,j}$ are defined as in Section~\ref{sec:prelim}. A position   $(r_1,c_1)\in[M]^2$  represents a symbol of $\hat{\mathcal B}$. It is unrelated to the grid coordinates $(r,c)\in[K_1]\times[K_2]$ which represent a cell of nodes in the network.

\textbf{Symbol-to-chunk assignment.} Illustrated in Fig.~\ref{fig:enc}, the object is divided into $K_2$ equally-sized chunks (assuming $K_2 | M$).  Chunk $c\in[K_2]$ consists of the following data columns
$\mathcal{J}_c=\Big\{(c-1)\tfrac{M}{K_2}+1,\ \dots,\ c\,\tfrac{M}{K_2}\Big\}$,
taken in all $\kappa$ layers and all $M$ rows. Symbol $(r_1,c_1)$ therefore belongs to chunk
$c=\Big\lceil c_1\,\tfrac{K_2}{M}\Big\rceil$.
We write $X_i^{(c)}$ for layer $i$ of $\mathcal{J}_c$, an $M\times(M/K_2)$ matrix. The chunk size is $|X^{(c)}|=\sigma/K_2$, where $\sigma $ is the size of the extended object $\hat{\mathcal B}$. Chunk $c$ is stored by column $c$ according to the PUT protocol (below).  Because a chunk contains whole data columns,   each being polynomially committed, any node that holds a coded version of a whole data column can compute KZG openings for it on demand (Section~\ref{sec:prelim}).

\subsection{PUT Protocol}
\label{sec:put}

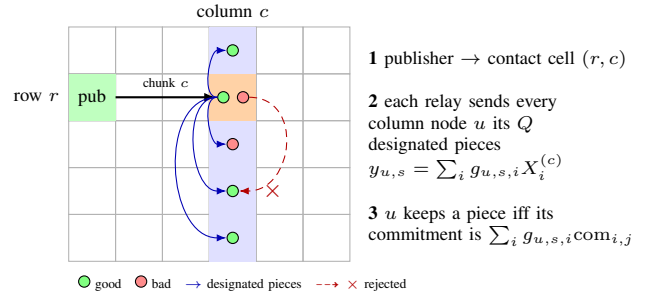
\begin{figure}[t]
\centering
\begin{tikzpicture}[scale=0.62,every node/.style={font=\scriptsize}]
\foreach \i in {0,...,4}{\foreach \j in {0,...,5}{\draw[gray!60] (\j,\i) rectangle ++(1,1);}}
\fill[blue!12] (3,0) rectangle (4,5);
\fill[orange!35] (3,3) rectangle (4,4);
\fill[green!25] (0,3) rectangle (1,4);
\node at (0.5,3.5) {pub};
\node[above] at (3.5,5) {column $c$};
\node[left] at (0,3.5) {row $r$};
\draw[-{Latex[length=1.6mm]},thick] (1,3.5) -- node[above]{\tiny chunk $c$} (3.15,3.5);
\node[circle,draw,fill=green!50,inner sep=1.5pt] (rg) at (3.3,3.5) {};
\node[circle,draw,fill=red!45,inner sep=1.5pt] (rb) at (3.72,3.5) {};
\node[circle,draw,fill=green!50,inner sep=1.5pt] (u1) at (3.5,4.5) {};
\node[circle,draw,fill=red!45,inner sep=1.5pt] (u2) at (3.5,2.5) {};
\node[circle,draw,fill=green!50,inner sep=1.5pt] (u3) at (3.5,1.5) {};
\node[circle,draw,fill=green!50,inner sep=1.5pt] (u4) at (3.5,0.5) {};
\foreach \u/\l in {u1/0.9,u2/0.9,u3/1.0,u4/1.1}{\draw[-{Latex[length=1.3mm]},blue!70!black] (rg.west) to[out=180,in=180,looseness=\l] (\u.west);}
\draw[-{Latex[length=1.3mm]},red!70!black,densely dashed] (rb.east) to[out=0,in=0,looseness=1.5] (u3.east);
\node[red!70!black,font=\small] at (4.35,1.5) {$\times$};
\node[right,align=left] at (6.2,4.3) {\textbf{1} publisher $\to$ contact cell $(r,c)$};
\node[right,align=left] at (6.2,2.65) {\textbf{2} each relay sends every\\column node $u$ its $Q$\\designated pieces\\$y_{u,s}=\sum_i g_{u,s,i}X_i^{(c)}$};
\node[right,align=left] at (6.2,0.75) {\textbf{3} $u$ keeps a piece iff its\\commitment is $\sum_i g_{u,s,i}\mathrm{com}_{i,j}$};
\node[font=\tiny,anchor=north west] at (0,-0.1) {\tikz\node[circle,draw,fill=green!50,inner sep=1.5pt]{}; good\quad \tikz\node[circle,draw,fill=red!45,inner sep=1.5pt]{}; bad\quad \textcolor{blue!70!black}{$\rightarrow$} designated pieces\quad \textcolor{red!70!black}{$\dashrightarrow\times$} rejected};
\end{tikzpicture}
\caption{PUT path: Every relay in contact cell $(r,c)$ sends each column node exactly its $Q$ designated pieces ($Q/\kappa$ of the chunk) instead of the full chunk; one good relay suffices. A piece with any other coefficients fails the commitment check and is dropped.}
\label{fig:put}
\end{figure}

PUT publishes an object $\hat{\mathcal B}$. Suppose that the publisher is on grid row $r$. It disseminates the $K_2$ chunks of $\hat{\mathcal B}$ in parallel, chunk $c$ into column $c$. Each node of column $c$ stores $Q\in [\kappa]$ \emph{coded pieces}, each a random linear combination of the chunk's $\kappa$ layers. The steps for putting each chunk $c\in[K_2]$ are as follows (illustrated in Fig. \ref{fig:put}).
\begin{enumerate}[leftmargin=*]
\item \emph{Row hop:} The publisher announces the commitments $\mathrm{com}_{i,j}$ and broadcasts chunk $c$ (all $\kappa$ layers) to all nodes in contact cell $(r,c)$; these nodes are the \emph{relays} into column $c$. If the contact cell has no good node, chunk $c$ is considered failed. TCDA does not detect or repair this: such misses are allowed by the threshold of $\delta$-robustness (Definition~\ref{def:robust}).
\item \emph{Column hop:} Each relay in contact cell $(r,c)$ computes, for every node $u$ in column $c$, $Q$ random coded pieces $\{y_{u,s}\}_{s \in [Q]}$  using {designated} coefficients for $u$, 
\[
g_{u,s}=H(\mathrm{com},c,u,s)\in\F^\kappa,\qquad y_{u,s}=\sum_{i=1}^{\kappa}g_{u,s,i}\,X_i^{(c)}
\]
and sends them to $u$. The relay itself keeps its own $Q$ designated pieces. 
\item \emph{Retention:} Each recipient node $u$ recomputes the commitment of each received coded data column (one multi-scalar multiplication of size $M$) and keeps $y_{u,s}$ iff it matches $C_{u,s,j}=\sum_i g_{u,s,i}\,\mathrm{com}_{i,j}$ for all $j\in\mathcal{J}_c$. Node $u$ ignores subsequent pieces after receiving its $Q$ valid pieces.
\end{enumerate}
Because a coded data column is stored wholly, no opening proofs are transmitted or stored at PUT time. They are computed on demand at GET time (Section~\ref{sec:prelim}).

\textbf{Why designated coefficients.} If relays chose the coding coefficients by themselves, one bad relay that reaches the column first could broadcast the same $Q$ pieces to every node. Each piece would pass verification, yet the good nodes would jointly hold rank at most $Q<\kappa$, and the chunk would be lost. TCDA therefore fixes each node's coefficients in advance by the random oracle, so exactly one piece is acceptable per slot and any other is rejected. For node $u$ of column $c$ and slot $s\in[Q]$,  since the Reed-Solomon code is linear, every data column of $y_{u,s}$ is again a Reed-Solomon codeword, with commitment $C_{u,s,j}=\sum_i g_{u,s,i}\,\mathrm{com}_{i,j}$ for $j\in\mathcal{J}_c$. Anyone can compute these commitments from public data alone, which is what the retention check of step~3 uses.

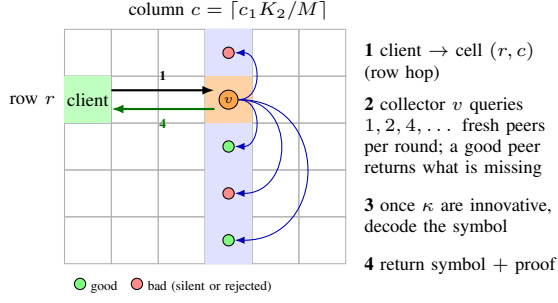
\begin{figure}[t]
\centering
\begin{tikzpicture}[scale=0.62,every node/.style={font=\scriptsize}]
\foreach \i in {0,...,4}{\foreach \j in {0,...,5}{\draw[gray!60] (\j,\i) rectangle ++(1,1);}}
\fill[blue!12] (3,0) rectangle (4,5);
\fill[orange!35] (3,3) rectangle (4,4);
\fill[green!25] (0,3) rectangle (1,4);
\node at (0.5,3.5) {client};
\node[above] at (3.5,5) {column $c=\lceil c_1K_2/M\rceil$};
\node[left] at (0,3.5) {row $r$};
\draw[-{Latex[length=1.6mm]},thick] (1,3.7) -- node[above]{\tiny \textbf{1}} (3.2,3.7);
\draw[-{Latex[length=1.6mm]},thick,green!45!black] (3.2,3.3) -- node[below]{\tiny \textbf{4}} (1,3.3);
\node[circle,draw,fill=orange!70,inner sep=1pt] (v) at (3.5,3.5) {\tiny $v$};
\node[circle,draw,fill=red!45,inner sep=1.5pt] (p1) at (3.5,4.5) {};
\node[circle,draw,fill=green!50,inner sep=1.5pt] (p2) at (3.5,2.5) {};
\node[circle,draw,fill=red!45,inner sep=1.5pt] (p3) at (3.5,1.5) {};
\node[circle,draw,fill=green!50,inner sep=1.5pt] (p4) at (3.5,0.5) {};
\draw[-{Latex[length=1.3mm]},blue!70!black] (v.east) to[out=0,in=0,looseness=1.4] (p1.east);
\draw[-{Latex[length=1.3mm]},blue!70!black] (v.east) to[out=0,in=0,looseness=1.4] (p2.east);
\draw[-{Latex[length=1.3mm]},blue!70!black] (v.east) to[out=0,in=0,looseness=1.6] (p3.east);
\draw[-{Latex[length=1.3mm]},blue!70!black] (v.east) to[out=0,in=0,looseness=1.7] (p4.east);
\node[right,align=left] at (6.2,4.3) {\textbf{1} client $\to$ cell $(r,c)$\\(row hop)};
\node[right,align=left] at (6.2,2.7) {\textbf{2} collector $v$ queries\\$1,2,4,\dots$ fresh peers\\per round; a good peer\\returns what is missing};
\node[right,align=left] at (6.2,1.0) {\textbf{3} once $\kappa$ are innovative,\\decode the symbol};
\node[right,align=left] at (6.2,-0.05) {\textbf{4} return symbol $+$ proof};
\node[font=\tiny,anchor=north west] at (0,-0.1) {\tikz\node[circle,draw,fill=green!50,inner sep=1.5pt]{}; good\quad \tikz\node[circle,draw,fill=red!45,inner sep=1.5pt]{}; bad (silent or rejected)};
\end{tikzpicture}
\caption{GET path for symbol $(r_1,c_1)$. A good collector $v$ in cell $(r,c)$ starts with its own $Q$ coordinates and queries fresh column peers in rounds of $1,2,4,\dots$ until $\lceil\kappa/Q\rceil-1$ good peers have supplied the rest. With $Q=\kappa$ the column hop is skipped.}
\label{fig:get}
\end{figure}

\subsection{GET Protocol}
GET samples a symbol at some random position $(r_1,c_1) \in [M]^2$. Suppose that the sampling client is on grid row $r$. It computes the symbol's chunk $c=\lceil c_1K_2/M\rceil \in [K_2]$, so it knows to find the symbol in column $c$. The data row $r_1$ plays no part in routing; it only selects which coordinates are decoded. GET has the following steps (illustrated in Fig.~\ref{fig:get}).
\begin{enumerate}[leftmargin=*]
\item \emph{Row hop:} The client queries every node in the contact cell $(r,c)$ for column $c$. Each good node of the cell acts as a \emph{collector} and starts with its own $Q$ coordinates $y_{v,s}[r_1,c_1]$ for the symbol. If $Q=\kappa$, this collector already holds $\kappa$ independent coordinates; it decodes and returns the symbol to the client. Else, the next step.  
\item \emph{Column hop:} Each collector queries its column peers in incremental rounds. In round $t=1,2,\dots$, it sends a request to $2^{t-1}$ peers not yet asked, chosen uniformly at random, for $\min\{Q,\kappa-k\}$ of their coordinates $y_{u,s}[r_1,c_1]$, where $k$ is the number of innovative coordinates it holds at the start of the round. A node $u$ returns them, each with a KZG opening against $C_{u,s,c_1}$ that $u$ computes itself. The collector stops querying the column after the round in which it has obtained $\kappa$ linearly-independent pieces.
\item \emph{Decode and reply:} The collector reconstructs the symbol given these $\kappa$   pieces, derives their openings by the same linear combination of the received proofs (KZG homomorphism), and returns the symbol with one aggregated proof. The client verifies it against the commitments $\mathrm{com}_{i,c_1}$.
\end{enumerate}

\subsection{JOIN Protocol}
\label{sec:join}
Nodes leave and join continually. Node departures would erode the number of good pieces.  Therefore, when a new node joins (returning is considered new), it counts toward the column's good nodes only once it stores its designated pieces for every object still in the retention window. Syncing these pieces is therefore what keeps TCDA correct under churn. A new node $u$  syncs each object in the following steps.
\begin{enumerate}[leftmargin=*]
\item \emph{Placement:} The random oracle places $u$ in cell $(r,c)$, and a bootstrap node connects it to its row and column peers.
\item \emph{Fetch:} Node $u$ queries its column peers in random order for their  pieces, each verified against its public commitment, until receiving $\kappa$ pieces with linearly-independent coefficient vectors.  Since each piece is $1/\kappa$ of the chunk, $u$ downloads one chunk's worth of data per object.
\item \emph{Derive:} For each slot $s\in[Q]$, node $u$ writes its designated vector $g_{u,s}$ as a combination $\sum_j\lambda_j g_j$ of the fetched coefficient vectors, sets $y_{u,s}=\sum_j\lambda_j y_j$, verifies it against $C_{u,s,\cdot}$, and discards the fetched pieces.
\end{enumerate}
Fetch succeeds whenever column $c$ holds $\lceil\kappa/Q\rceil$ synced good nodes. Until successfully synced, $u$ retries and does not count toward the column's good nodes. 

\subsection{TUNE Protocol}
\label{sec:tune}
The retention depth $Q$ is a public protocol parameter per data type. It is set by the design rule (later proposed in Section~\ref{sec:design}). The feasibility range of $Q$ is bounded by $\delta$-robustness from below and the PUT and storage budgets from above. An adjustment of $Q$ is needed when:
\begin{itemize}[leftmargin=*]
\item \emph{Safety (raise):} the lower bound on number of good nodes $\Np$ falls and $Q$ no longer meets $\delta$-robustness.
\item \emph{Deadline (lower):} network size $N$ grows, pushing the PUT load over the budget; if $Q$ cannot decrease further, the grid must be widened.
\item \emph{Margin (raise, optional):} the budgets leave slack, so a larger $Q$ widens the safety margin against good-node departures.
\item \emph{Cost (lower, optional):} good-node growth lets a smaller $Q$ meet $\delta$-robustness and so lowering $Q$  cuts PUT and storage.
\end{itemize}

Upon a change in retention depth $Q$, nodes adjust stored objects without republishing, simply as follows:
\begin{enumerate}[leftmargin=*]
\item \emph{Raise $Q$ to $Q'>Q$.} Every node runs the Fetch and Derive steps of JOIN for the new slots $s=Q+1,\dots,Q'$ only.
\item \emph{Lower $Q$ to $Q'<Q$.} Every node deletes its slots $s=Q'+1,\dots,Q$. No communication is needed.
\end{enumerate}

\section{Robustness Conditions}
\label{sec:security}

A chunk $c$ is \emph{retrievable} if, once it has entered column $c$ through a good relay, (i)~the pieces held by the good nodes of column $c$ can determine chunk $c$, and (ii)~every GET for a symbol of chunk $c$ that reaches a good node of column $c$ returns the correct symbol.

\begin{lemma}[Chunk retrievability]
\label{lem:retrieve}
Chunk $c$ is retrievable iff
\[
\Ncp\ \ge\ \hQ \triangleq \lceil\kappa/Q\rceil,
\]
where the ``if'' direction holds except with probability at most $\binom{\Ncp}{\hQ}/(q-1)$, which is negligible for $q\approx2^{255}$ (the field size). Moreover, every GET then completes within $\lceil\log_2(N_c-\Ncp+\hQ)\rceil$ rounds of column queries.
\end{lemma}
\begin{proof}

\emph{Necessity.} Suppose $\Ncp<\hQ$. We have only $\Ncp Q<\kappa$ coefficient vectors of the good nodes, not enough to recover chunk $c$.

\emph{Sufficiency.} Suppose $\Ncp\ge\hQ$. A good node keeps a piece only if it matches the public commitment, so by KZG binding each good node holds exactly its $Q$ designated pieces. Their coefficient vectors are random-oracle outputs of the fixed node identities, uniform in $\F^\kappa$ and beyond the adversary's control. Any $\hQ$ good nodes thus hold $\hQ Q\ge\kappa$ pieces, which have rank $\kappa$ except with probability $<1/(q-1)$. By a union bound over the $\binom{\Ncp}{\hQ}$ such subsets, every one of them can reconstruct chunk $c$ except with probability $<\binom{\Ncp}{\hQ}/(q-1)$. The returned symbol is correct, since every coordinate is checked against the commitment: a bad responder can only send coordinates that are rejected, or valid ones that do no harm.

After $\lceil\log_2(N_c-\Ncp+\hQ)\rceil$ doubling rounds, the collector has queried at least $(N_c-\Ncp+\hQ-1)$ peers, hence $(\hQ-1)$ good ones, and holds rank $\kappa$, whatever the bad peers send or withhold. Decoding and proof derivation follow from linearity and KZG homomorphism.

\end{proof}

Next, we will derive the sampling failure rate. A GET for a symbol fails due to access failure or decoding failure. Access failure means that we cannot get to the custody column, happening if the publisher's contact cell   or the client's contact cell has no good node. Decoding fails if the custody column $c$ holds fewer than $\hQ$ good nodes, $\Ncp < \hQ$ (Lemma \ref{lem:retrieve}). Fix a good publisher and a good client in different rows, the worst placement, and a column $c$ that is neither's; the other $N^+$ good nodes are placed uniformly.

\begin{lemma}[Access failure]
\label{lem:p1}
The publisher's or the client's contact cell holds no good node with probability
\[
P_1:= 2\Big(1-\frac{1}{K_1K_2}\Big)^{\Np}-\Big(1-\frac{2}{K_1K_2}\Big)^{\Np}.
\]
\end{lemma}
\begin{proof}
Each good node misses a given cell with probability $1-1/(K_1K_2)$ and both cells with $1-2/(K_1K_2)$, independently of the others. Inclusion--exclusion gives $P_1$.
\end{proof}

\begin{lemma}[Decoding failure]
\label{lem:p2}
Let $\pi_n=\binom{\Np}{n}K_2^{-n}(1-1/K_2)^{\Np-n}$ and $f_n=1-2(1-1/K_1)^n+(1-2/K_1)^n$. Given that both contact cells hold a good node, the custody column holds fewer than $\hQ$ good nodes with probability
\[
P_2  =\frac{1}{1-P_1}\sum_{n=0}^{\hQ-1}\pi_nf_n .
\]
\end{lemma}
\begin{proof}
Given uniform placement of the good nodes, the good count in the column is binomially distributed,   $\Ncp\sim\mathrm{Bin}(\Np,1/K_2)$, and so $\Pr[\Ncp=n]=\pi_n$. Given $\Ncp=n$, the $n$ good nodes of the column are uniformly placed into its $K_1$ cells, so both contact cells are hit with probability $f_n$, by inclusion-exclusion. Decoding fails for every $n<\hQ$ (Lemma \ref{lem:retrieve}), hence probability $\sum_{n=0}^{\hQ-1}\pi_nf_n$ for decoding failure. Conditioned on a successful access (probability $1-P_1$ from Lemma \ref{lem:p1}), we obtain $P_2$. 
\end{proof}

Because a GET fails due to failure of either access   or decoding, the probability of sampling failure is $P_1+(1-P_1)P_2$. Hence, we obtain the result below.
\begin{corollary}[Sample failure]
\label{cor:sample}
Against the worst-case adversary, and up to the negligible rank failure of Lemma~\ref{lem:retrieve}, a sampling fails with probability exactly
\begin{align*}
P  = \underbrace{2\Big(1-\frac{1}{K_1K_2}\Big)^{\Np}-\Big(1-\frac{2}{K_1K_2}\Big)^{\Np} +\sum_{n=0}^{\hQ-1}\pi_n f_n}_{\triangleq p_{\mathrm{fail}}(\Np,K_1,K_2,Q)}
\end{align*}
\end{corollary}

This probability increases in $K_1$ and decreases in $Q$. The following inequality is useful for closed-form bounding:  
\begin{align}
&P \le 2\,e^{-\Np/(K_1K_2)}  +  \nonumber \\
&\begin{cases}
0 & \text{if~} h_Q = 1\\
e^{-\Nbarp}\Big(\frac{e\,\Nbarp}{\hQ-1}\Big)^{\hQ-1} & \text{if ~} h_Q\ge2,~ \bar N^+_c>h_Q-1\\
1 & \text{otherwise.}
\end{cases}
\label{eq:Pbound}
\end{align}

TCDA is  $\delta$-robust iff $P \le \delta$. We thus immediately obtain the following condition for TCDA to safely bootstrap.
\begin{theorem}[Bootstrap condition]
\label{thm:bootstrap}
Given a design  $(K_1,K_2,Q)$, TCDA  is $\delta$-robust iff the network has at least this many good nodes:
\[
 \Np \ge \Np_\delta(Q) =\min\{\Np : p_{\mathrm{fail}}(\Np,K_1,K_2,Q)\le\delta\}.
\]
The minimum $\Np_\delta(Q)$ is non-increasing in $Q$ and non-decreasing in $K_1$. Furthermore, we can upper-bound it:
\begin{align}
\Np_\delta(Q) \le
\begin{cases}
K_1K_2\ln(2/\delta) & \text{if~} h_Q = 1\\
\max\big\{K_1K_2\ln(4/\delta),\ K_2\,x_Q\big\} & \text{otherwise.}
\end{cases}
\label{eq:NpdeltaQ}
\end{align}
where    $x_Q$ is the root $x > \hQ-1$ of equation 
\[
x-(\hQ-1)\ln\frac{ex}{\hQ-1}=L :=\ln(2/\delta).
\] 
This $x_Q$ satisfies
\begin{align}
h_Q-1 < x_Q \le\ \Big(\sqrt{L/2}+\sqrt{\hQ-1+L/2}\Big)^2.
\label{eq:xQ}
\end{align}
\end{theorem}
\begin{proof}
The bootstrap condition is obvious and the monotonicity in $K_1$ and $Q$ is easy to derive. Let us focus on the closed-form inequality. 
By Inequality ~\eqref{eq:Pbound}, it suffices that its closed-form bound be at most $\delta$. 

For $\hQ=1$, letting $2e^{-\Np/(K_1K_2)}\le\delta$, we have $\Np\ge K_1K_2\ln(2/\delta)$. 
For $\hQ\ge2$, bounding each term by $\delta/2$, we have
\begin{itemize}
\item The first term: 
$2e^{-\Np/(K_1K_2)}\le\delta/2 \Rightarrow \Np\ge K_1K_2\ln(4/\delta)$.
\item The second term:
\begin{align*}
\underbrace{e^{-\Nbarp}(e\Nbarp/(\hQ-1))^{\hQ-1}}_{\triangleq g(\Nbarp) }\le \underbrace{\delta/2}_{ = g(x_Q)}.
\end{align*}
Since $g$ decreases for $\Nbarp>\hQ-1$ and $x_Q > h_Q-1$, we get $\Nbarp\ge x_Q \Rightarrow \Np \ge K_2 x_Q$. 
\end{itemize}
For the upper bound on $x_Q$, let $\mu_0$ be the value at which $\hQ-1=(1-t)\mu_0$ with $t=\sqrt{2L/\mu_0}$, i.e., $\sqrt{\mu_0}=\sqrt{L/2}+\sqrt{\hQ-1+L/2}$. Since $(1-t)\ln(1-t)\ge-t+t^2/2$, $g(\mu_0)=e^{-\mu_0[t+(1-t)\ln(1-t)]}\le e^{-t^2\mu_0/2}=\delta/2$, so $x_Q\le\mu_0$. 
\end{proof}

\begin{corollary}[Safe range of the knob]
\label{cor:qdelta}
Consider a network with $\Np$ good nodes placed uniformly. Let $\Nbarp=\Np/K_2$ and $L=\ln(2/\delta)$.
\begin{enumerate}[leftmargin=*,label=(\roman*)]
\item \emph{Tunability.} TCDA is $\delta$-robust iff 
\[
Q\ge Q_\delta:=\min\{Q\in[\kappa]:p_{\mathrm{fail}}(\Np,K_1,K_2,Q)\le\delta\}.
\]
\item \emph{Existence.} $Q_\delta$ exists iff $\Np\ge \Np_\delta(\kappa)$, the bootstrap of full replication, and
\[
(K_1K_2-1)\ln\frac{1}{1-\sqrt{1-\delta}}\ \le\ \Np_\delta(\kappa)\ \le\ K_1K_2\ln \frac{2}{\delta}.
\]
\item \emph{Closed form.} If $\Np\ge K_1K_2\ln(4/\delta)$ and $\Nbarp\ge2L$, then
\[
Q_\delta\ \le\ \big\lceil\kappa/h\big\rceil,\qquad h=\big\lfloor\Nbarp-\sqrt{2L\Nbarp}\big\rfloor+1 .
\]
\end{enumerate}
\end{corollary}
\begin{proof}
\emph{(i)} Since $p_{\mathrm{fail}}$ decreases in $Q$, the tunability range is an immediate consequence of the bootstrap condition in Theorem \ref{thm:bootstrap}.

\emph{(ii)} Also, thanks to Theorem \ref{thm:bootstrap}, $\Np_\delta(Q)$ is non-increasing in $Q$ and so  $\Np_\delta(\kappa) \le \Np_\delta(Q) \le \Np_\delta(Q_\delta)$ for all $Q \in [Q_\delta, \kappa]$. Therefore, $Q_\delta$ exists iff $\Np\ge\Np_\delta(\kappa)$. The upper bound of $\Np_\delta(\kappa)$ is according to Eq \eqref{eq:NpdeltaQ} by setting $Q =\kappa$ (i.e., $h_Q=1$). For the lower bound, according to Corollary \ref{cor:sample}, by setting $Q=\kappa$ and $\Np = \Np_\delta(\kappa)$, we have $p_{\mathrm{fail}} = P = 2p-p_2$ where $p=(1-1/(K_1K_2))^{\Np}$ and $p_2=(1-2/(K_1K_2))^{\Np}\le p^2$. Hence 
\[
p_{\mathrm{fail}}\le\delta \Rightarrow 1-(1-p)^2\le\delta \Rightarrow p\le\delta_{\mathrm c} := 1-\sqrt{1-\delta}.
\]
Applying inequality  $-\ln(1-x)\le x/(1-x)$, we obtain $\Np_\delta(\kappa) = \Np  \ge(K_1K_2-1)\ln(1/\delta_{\mathrm c})$.

\emph{(iii)} Let $Q=\lceil\kappa/h\rceil$. Then $\kappa/Q\le h$, so $\hQ\le h$ and so
\begin{align*}
\hQ-1 \le h-1 \le \Nbarp-\sqrt{2L\Nbarp}  \\
\Rightarrow \sqrt{\Nbarp}\ge\sqrt{L/2}+\sqrt{\hQ-1+L/2}.
\end{align*}
 By Eq \eqref{eq:xQ}, we have $x_Q\le\Nbarp$ if $\hQ\ge2$, and together with $\Np\ge K_1K_2\ln(4/\delta)$ this gives $\Np\ge\Np_\delta(Q)$; if $\hQ=1$, $\Np\ge K_1K_2\ln(2/\delta)$ suffices. Hence $Q_\delta\le Q=\lceil\kappa/h\rceil$.
\end{proof}

Together, Theorem~\ref{thm:bootstrap} and Corollary~\ref{cor:qdelta} show what the bootstrap is made of. Parameter $Q$ allows TCDA to tune priorities in terms of PUT, GET, and storage costs. Note that, at $Q=\kappa$, every node keeps all $\kappa$ pieces, the size of full replication, and TCDA has the same storage and dissemination protocol as that of RDA. However, our bootstrap requirement is less demanding because we allow for columns to be empty of good nodes, whereas RDA requires every column to contain one. 

\textbf{Example.} Take $(15, 32)$ grid, erasure-coding granularity $\kappa=64$, and robustness requirement $\delta = 1/4$. Table~\ref{tab:example} lists for each $Q$ the smallest number of good nodes, $\Np_\delta(Q)$,   that makes TCDA $\delta$-robust, computed exactly and from the closed form according to Theorem \ref{thm:bootstrap}, which shows a close approximation (at most $38\%$ over).

\begin{table}[h]
\centering
\caption{TCDA: bootstrap size for $(15, 32)$ grid, $\kappa=64$, $1/4$-robustness.}
\label{tab:example}
\scriptsize
\setlength{\tabcolsep}{5pt}
\begin{tabular}{@{}rrrr@{}}
\toprule
$Q$ & $\hQ$ & $\Np_\delta(Q)$ exact & closed form\\
\midrule
64 & 1 & 965 & 999\\
16 & 4 & 965 & 1331\\
8 & 8 & 965 & 1331\\
4 & 16 & 966 & 1331\\
2 & 32 & 1222 & 1401\\
1 & 64 & 2221 & 2580\\
\bottomrule
\end{tabular}
\end{table}

\begin{table}[h]
\centering
\caption{TCDA vs. RDA: on same $(15, 32)$ grid, $\kappa=64$,  same $\delta$-robustness.}
\label{tab:rda}
\scriptsize
\setlength{\tabcolsep}{5pt}
\begin{tabular}{@{}lrrcc@{}}
\toprule
$\epsilon$ & $\Np$ & $\delta_{\mathrm{RDA}}$ & RDA's $Q$ & TCDA's $Q$\\
\midrule
$1/4$ & 150 & 0.93 & 64 & 32--64\\
$10^{-3}$ & 327 & 0.76 & 64 & 22--64\\
$10^{-6}$ & 545 & 0.54 & 64 & 11--64\\
$10^{-9}$ & 762 & 0.37 & 64 & 7--64\\
\bottomrule
\end{tabular}
\end{table}

Table~\ref{tab:rda} compares TCDA to RDA. Given the same network sized by RDA under various failing rates ($\epsilon$) of enforcing its good-node assumption, each failing rate leading to a corresponding degree of robustness,  $\delta_{\mathrm{RDA}}$. For the same robustness, TCDA offers a free choice of $Q$ in the listed range, whereas RDA is fixed at $Q=\kappa$. Not only that, the larger $\Np$ (more good nodes), the more freedom to tune $Q$.

\section{Cost Model}
\label{sec:cost}

 Storage and PUT traffic scale with the object, so we measure them in units of its size $\sigma$. A GET returns one symbol whatever the object's size, so we measure it in units of the symbol size $\sigma_s=\sigma/M^2$. Let $\sigma_{\mathrm{pf}}$ be the size of a KZG proof ($48$\,B) and $\beta=\sigma_{\mathrm{pf}}/\sigma_s$.   The sync of JOIN costs the same in both TCDA and RDA; it does not depend on $Q$.

\subsection{Storage Cost}
Let $\Phi_{\mathrm{store}}$ be the amount of storage required at  a node per object. Objects are retained for an availability window of $W$ blocks, one object per block, so a node holds $W\Phi_{\mathrm{store}}$ at any time. A node keeps   $Q$ designated pieces, each a combination of the chunk's $\kappa$ layers and hence of size $\sigma/(\kappa K_2)$:
\[
\Phi_{\mathrm{store}}= \bigg( \frac{Q}{\kappa K_2} \bigg) \sigma .
\]

\subsection{PUT Cost}
Let $\Phi_{\mathrm{put}}$ be the amount of data a good relay sends in the column hop of PUT, per object. This is the per-node load that the attestation deadline caps. There are $K_2$ chunks per object, each published in a separate column. Suppose that the publisher is in row $r$. It sends these $K_2$ chunks to their respective cells in row $r$. For each chunk $c$, every good relay in cell $(r, c)$ sends $Q$ pieces to each of the $N_c$ nodes in column $c$, so a relay of column $c$ sends
\begin{align}
\Phi_{\mathrm{put}}(c)=\frac{Q\sigma}{\kappa  K_2} N_c.
\label{eq:PhiPut}
\end{align}
The total PUT traffic per object, the publisher's row hop included, is
\begin{align*}
\Phi^{\mathrm{net}}_{\mathrm{put}}(r)
&= \frac{N_r \sigma}{K_2}+ \sum_{c=1}^{K_2} N^+_{r,c}\, \Phi_{\mathrm{put}}(c) \\
&=\Big ( \frac{N_r}{K_2}+\frac{Q}{\kappa K_2}\sum_{c}N^+_{r,c}\,N_c\Big ) \sigma,
\end{align*}
where $N_r$ is the population of row $r$ and $N^+_{r,c}$ the good count of cell $(r,c)$. Assuming uniform distributions of good nodes and publishers in the network, its mean is
\begin{align*}
\Phi^{\mathrm{net}}_{\mathrm{put}} := \mathbb E [\Phi^{\mathrm{net}}_{\mathrm{put}}(r)]  \approx \Big ( \frac{N}{K_1K_2}+\frac{Q}{\kappa}\cdot\frac{\Np N}{K_1K_2^2}\Big) \sigma.
\end{align*}

\subsection{GET Cost}
\label{sec:getcost}
In GET, each collector node already holds $Q$ coordinates of the symbol. By Lemma~\ref{lem:retrieve}, it must obtain the remaining $(\kappa-Q)$ from
$m:=\hQ-1=\lceil\kappa/Q\rceil-1$
good column peers. These peers are queried in rounds, each round doubling the contact size: 1, 2, 4, ...

\textbf{Latency (successful GET). } The collector stops after $t=\lceil\log_2(X+1)\rceil$ rounds where $X$ is the position, in the collector's random order of its $N_c-1$ peers, of the $m$-th good one contacted. The expectation notation in the following is over randomness of $X$. Suppose that the symbol belongs to chunk $c$. We have
\begin{align}
\mathbb E[X]= \frac{m}{\eta_c},\qquad \eta_c=\frac{\Ncp}{N_c},
\label{eq:X}
\end{align}
the negative hypergeometric mean. Counting the one round-trip from the sampling client to its contact cell, the latency of a successful GET attempt is quantified by the following number of round-trips
\begin{align*}
\Lambda^{\mathrm{succ}}_{\mathrm{get}}(c) := \mathbb E[1+t] \le 2+\log_2\Big(\frac{m}{\eta_c}+1\Big).
\end{align*}
The adversary can worsen the good fraction $\eta_c$ of a column by adding arbitrarily many bad nodes, but the impact on latency is only logarithmic because
\begin{align*}
t \le \lceil\log_2(N_c-\Ncp+\hQ)\rceil.
\end{align*}
The collector has then queried $\Gamma_{\mathrm{get}}=2^t-1<2X$ column peers, so $\mathbb E[\Gamma_{\mathrm{get}}]<2m/\eta_c$.

\textbf{Traffic (successful GET). }  Let $\Phi^{\mathrm{succ}}_{\mathrm{get}}$ denote the expected amount of data all good nodes send for one successful GET attempt. Per collector, the good peers collectively send at least the $(\kappa-Q)$ coordinates.  Each coordinate carries a proof, and the collector returns the symbol with one aggregate proof. Summing over the $N^+_{r,c}$ collectors, we have a lower bound:
\begin{align*}
\Phi^{\mathrm{succ}}_{\mathrm{get}}(r, c) \ge N^+_{r,c}\Big[(\kappa-Q)\Big(\frac1\kappa+\beta\Big)+(1+\beta)\Big]\sigma_s.
\end{align*}

Also, we know that each good column node returns at most $Q$ coordinates for the symbol. The number of good nodes contacted is at most $\Gamma_{\mathrm{get}}<2X$. Applying Eq \eqref{eq:X}, we obtain an upper bound:
\begin{align*}
 \Phi^{\mathrm{succ}}_{\mathrm{get}}(r, c) < N^+_{r,c}\Big[\frac{2mQ}{\eta_c}\Big(\frac1\kappa+\beta\Big)+(1+\beta)\Big]\sigma_s.
\end{align*}

\textbf{Latency and traffic per GET, failures included. } The above formulation applies to a successful GET attempt. In general, an attempt can fail, which incurs wasted costs. A failure is  for one of two reasons: i) access failure: a contact cell holds no good node (probability $P_1$, Lemma~\ref{lem:p1}) or ii) decoding failure: access is successful but the column is not decodable (probability $P_2$, Lemma~\ref{lem:p2}). 
Let $\Lambda_1,\Phi_1$ and $\Lambda_2,\Phi_2$ be the latency-traffic costs of an attempt that fails due to these reasons, respectively.
Then, the expected latency and traffic costs to get a verified symbol, in terms of round-trips, are
\begin{align}
\Lambda_{\mathrm{get}}=  \Lambda^{\mathrm{succ}}_{\mathrm{get}} +\frac{P_1\,\Lambda_1}{(1-P_1)(1-P_2)}+\frac{P_2\,\Lambda_2}{1-P_2}\\
\Phi_{\mathrm{get}}=  \Phi^{\mathrm{succ}}_{\mathrm{get}} +\frac{P_1\,\Phi_1}{(1-P_1)(1-P_2)}+\frac{P_2\,\Phi_2}{1-P_2}
\end{align}
Here, the row/column index of $\Phi, \Lambda$ is omitted for clarity.
\begin{itemize}
\item \emph{Access failure:} If the client's contact cell has no good node, the sampling query stops there. If the publisher's contact cell has no good node, the chunk never entered the column, and as such no good node of the client's contact cell holds any piece. Therefore, $\Lambda_1=1$ and $\Phi_1=0$.
\item \emph{Decoding failure}. Each collector doubles its contact size until it has queried its whole column; hence $\Lambda_2=1+\lceil\log_2N_c\rceil$. The collector has $\Ncp-1\le\hQ-2$ good peers, each sending at most $Q$ coordinates, with no reply to the client. So 
\begin{align*}
\Phi_2
&\le N^+_{r,c}(\Ncp-1)\,Q\Big(\frac1\kappa+\beta\Big)\sigma_s\\
&<N^+_{r,c}(\kappa-Q)\Big(\frac1\kappa+\beta\Big)\sigma_s.
\end{align*}
\end{itemize}
Substituting the above, TCDA needs
\begin{align*}
\Lambda_{\mathrm{get}}&= \Lambda^{\mathrm{succ}}_{\mathrm{get}} + \frac{P_1}{(1-P_1)(1-P_2)} +\frac{P_2}{1-P_2}\big(1+\lceil\log_2N_c\rceil\big)\\
\Phi_{\mathrm{get}}&= \Phi^{\mathrm{succ}}_{\mathrm{get}} +\frac{P_2}{1-P_2}\,\Phi_2.
\end{align*}

\begin{remark}
The formulation of the costs is summarized in Table~\ref{tab:cost}. Compared to RDA, storage and PUT costs fall by $Q/\kappa$. In exchange, every GET takes a second hop, of about $\log_2(m/\eta_c)+1$ round-trips, fewer as $Q$ grows, and moves at least $(\kappa-Q)/\kappa$ of a symbol more.
\end{remark}

\begin{table*}[t]
\centering
\caption{Cost model ($\Phi^{\mathrm{net}}_{\mathrm{put}}$ at mean populations). $m=\lceil\kappa/Q\rceil-1$, $\eta_c=\Ncp/N_c$, $\beta=\sigma_{\mathrm{pf}}/\sigma_s$; $N^+_{r,c}$, the good count of the client's contact cell, is common to both designs and about $\Np/(K_1K_2)$ on a typical cell; $P_1$ and $P_2$ are given by Lemmas~\ref{lem:p1} and~\ref{lem:p2}. At $Q=\kappa$, TCDA coincides with RDA. \colorbox{hl}{Shaded in green}\,: the costs in the design problem~\eqref{eq:opt} of Section~\ref{sec:design}, namely the storage and PUT budgets and the GET latency to minimize.}
\label{tab:cost}
\small
\setlength{\tabcolsep}{6pt}
\begin{tabular}{@{}llcc@{}}
\toprule
 & Cost & RDA & TCDA$(Q<\kappa)$ \\
\midrule
Per object & \cellcolor{hl}$\Phi_{\mathrm{store}}/\sigma$ & \cellcolor{hl}$1/K_2$ & \cellcolor{hl}$\frac{Q}{\kappa}/K_2$ \\
 & \cellcolor{hl}$\Phi_{\mathrm{put}}/\sigma$ & \cellcolor{hl}$N_c/K_2$ & \cellcolor{hl}$\frac{Q}{\kappa}N_c/K_2$\\
 & $\Phi^{\mathrm{net}}_{\mathrm{put}}/\sigma$ (column hop) & $\Np N/(K_1K_2^2)$ & $\frac{Q}{\kappa}\Np N/(K_1K_2^2)$\\
\midrule
Per successful attempt & $\Lambda^{\mathrm{succ}}_{\mathrm{get}}$ (round-trips) & $1$ & $\le2+\log_2(\frac{m}{\eta_c}+1)$\\
 & $\mathbb E[\Gamma_{\mathrm{get}}]$ (column queries) & $0$ & $<2m/\eta_c$\\
 & $\Phi^{\mathrm{succ}}_{\mathrm{get}}/\sigma_s$ & $N^+_{r,c}(1+\beta)$ & $\ge N^+_{r,c}\big[(\kappa-Q)(\frac1\kappa+\beta)+1+\beta\big]$\\
\midrule
Per sample & \cellcolor{hl}$\Lambda_{\mathrm{get}}$ (round-trips) & \cellcolor{hl}$\dfrac{1}{1-P_1}$ & \cellcolor{hl}$\Lambda^{\mathrm{succ}}_{\mathrm{get}}+\dfrac{P_1}{(1-P_1)(1-P_2)}+\dfrac{P_2}{1-P_2}\big(1+\lceil\log_2N_c\rceil\big)$\\[2ex]
 & $\Phi_{\mathrm{get}}/\sigma_s$ & $N^+_{r,c}(1+\beta)$ & $\le\Phi^{\mathrm{succ}}_{\mathrm{get}}/\sigma_s+\dfrac{P_2}{1-P_2}\,N^+_{r,c}(\Ncp-1)\,Q\big(\frac1\kappa+\beta\big)$\\
\bottomrule
\end{tabular}
\end{table*}

\section{Optimal Design}
\label{sec:design}

In blockchain DAS, the costs of storage, PUT, and GET weigh differently. Storage must fit the disk space a node gives to DAS, which holds the objects of the last $W$ blocks, so $W\Phi_{\mathrm{store}}\le B_{\mathrm S}$, where $B_{\mathrm S}$ is the storage budget per node. The PUT protocol sits on the path to finality: every relay must finish the column hop of each block before validators attest, so the per-relay PUT cost should be capped by a budget, $\Phi_{\mathrm{put}} \le B_{\mathrm P}$: the uplink rate times the part of the attestation deadline left once the publisher has reached the relays.  The GET protocol has no such cap: clients sample continually, and every round-trip saved shortens every sample. However, we want to make GET as fast as possible, i.e., min $\Lambda_{\mathrm{get}}$.

\textbf{Example (Ethereum).} A slot lasts $12$\,s, validators attest about $4$\,s into it~\cite{paradigm-slot}, and an attesting node is recommended $25$\,Mbit/s of uplink~\cite{eip7870}. A publisher with a $1$\,Gbit/s uplink reaches the relays of every column within about $0.87$\,s, which leaves $B_{\mathrm P}=9.3$\,MiB per relay per object. Blobs are retained for $W=131{,}072$ blocks, about $18$ days~\cite{eip4844}, and we give DAS $B_{\mathrm S}=32$\,GiB per node, half of what PeerDAS \cite{peerdas} stores per node, i.e., $B_{\mathrm S}/W=0.25$\,MiB per object. For $\sigma=32$\,MiB, the budgets are $0.29\sigma$ and $\sigma/128$. On the $(15,32)$ grid with $965$ good nodes and as many bad ones, a column holds $60$ nodes on average and up to $79$ at the $99$th percentile, so RDA's relay load reaches $\Phi_{\mathrm{put}}=79\sigma/K_2=2.47\,\sigma$, $8.5$ times the PUT budget, and each node stores $\sigma/K_2$, four times the storage budget. RDA thus fails both budgets.

\subsection{Design Problem}
\label{sec:designproblem}

Once PUT and storage fit their budgets, the goal is to achieve the fastest GET. All this is subject to $\delta$-robustness. Speaking of constraints, PUT is more important than storage. Storage can be cheap today, but PUT must meet the attestation deadline. So, optionally, we can drop the storage constraint to have more search space to minimize GET, but PUT is a hard constraint. 

The costs, as defined in the previous sections, depend on the network size $N$, good and bad, which is not known in the design phase. If we allow $N$ to grow any large,  the adversary can add arbitrarily many Sybil nodes to increase the costs boundlessly. Therefore, we must set a maximum network size, $N \le N_{\max}$. This worst-case sizing does not affect robustness: $\delta$-robustness depends only on the good nodes, $\Np\ge\Np_\delta(Q)$ (Theorem~\ref{thm:bootstrap}). If the network grows beyond $N_{\max}$ nodes, only the PUT load and GET latency go beyond the budget, but it remains as robust.

The inputs to our design problem are: lower bound $\Np$ on the good nodes, upper bound $N_{\max}$ on the total population, robustness $\delta$, object parameters ($\sigma$, $M$, $\kappa$),  budgets ($B_{\mathrm P}$, $B_{\mathrm S}$, $W$), and peer-degree bound $d_{\max}$. The design is the following optimization: 
\begin{align}
\min_{K_1,\,K_2,\;Q}\quad & \Lambda_{\mathrm{get}}(\Np,N_{\max};K_1,K_2,Q) \quad \text{(latency)}\label{eq:opt}\\
\text{s.t.}\quad & p_{\mathrm{fail}}(\Np;K_1,K_2,Q)\le\delta, \quad \text{(robustness)}\notag\\
& \Phi_{\mathrm{put}}(N_{\max};K_2,Q)\le B_{\mathrm P}, \quad \text{(deadline)}\notag\\
& W\,\Phi_{\mathrm{store}}(K_2,Q)\le B_{\mathrm S}, \quad \text{(storage)}\notag\\
& Q\in [\kappa],\quad K_2\mid M, \quad 1/K_1+1/K_2 \le d_{\max}. \notag
\end{align}
The last constraint is to keep a low overhead for a node to maintain peer connections in the network, about $N(1/K_1+1/K_2)$ of them. For example, $d_{\max}=0.05$ keeps the overhead below $5\%$ of the network size. Without this constraint, $K_1=1$ is optimal, i.e., a node needs to maintain connections with every other node, practically unreasonable.

The costs in the design problem are exactly formulated as follows:
\begin{itemize}
\item Columns differ in size, and the PUT must finish in every column, not only in a typical one, so we size the PUT on $\hat N_c$, the $(1-\epsilon_{\mathrm P})$-quantile of the column population $\mathrm{Bin}(N_{\max},1/K_2)$; we take $\epsilon_{\mathrm P}=0.01$ ($99\%$-quantile). 
\item The GET cost is a mean over samples, so it uses the typical column, $N_c=\Nbar=N_{\max}/K_2$ and $\Ncp=\Nbarp=\Np/K_2$, and $\eta_c=\eta=\Np/N_{\max}$. 
\end{itemize}
The constraints then are equivalent to
\begin{align}
\Np\ge\Np_\delta(Q),  \quad
\frac{Q}{\kappa}\,\frac{\hat N_c}{K_2}\,\sigma\le B_{\mathrm P}, \quad
 \frac{Q}{\kappa}\,\frac{\sigma}{K_2}\,W\le B_{\mathrm S} \label{eq:constraints},
\end{align}
and the objective is
\begin{align}
\Lambda_{\mathrm{get}}=\Lambda^{\mathrm{succ}}_{\mathrm{get}}+\frac{P_1}{(1-P_1)(1-P_2)}+\frac{P_2\big(1+\lceil\log_2\Nbar\rceil\big)}{1-P_2},
\label{eq:objective}
\end{align}
with $\Lambda^{\mathrm{succ}}_{\mathrm{get}}\le2+\log_2(m/\eta+1)$ at $Q<\kappa$ and $=1$ at $Q=\kappa$.

\subsection{Solution Approach}
\label{sec:solution}
Our solution is guided by the following results. First, on each grid, the optimal $Q$ has a closed form.

\begin{corollary}[Optimal retention depth $Q$]
\label{cor:window}
Fix the grid $(K_1,K_2)$. Problem~\eqref{eq:opt} is feasible iff 
\begin{align*}
Q_\delta\le Q_{\max} 
:=\min \bigg \{\kappa, 
\underbrace{\lfloor\kappa B_{\mathrm P}K_2/(\hat N_c\sigma)\rfloor}_{\triangleq Q_{\mathrm P}},
\underbrace{\lfloor\kappa B_{\mathrm S}K_2/(W\sigma)\rfloor}_{\triangleq Q_{\mathrm S}}
\bigg\}
\end{align*}
and then, $Q=Q_{\max}$ is the optimal.
\end{corollary}
\begin{proof}

$\delta$-robustness bounds $Q$ from below by $Q_\delta$ (Corollary~\ref{cor:qdelta}(i)). We need $Q \le Q_{\mathrm P}$  and $Q \le Q_{\mathrm S}$ because of the constraints involving $Q$ in Eq \eqref{eq:constraints}. Hence the feasible values on the grid form the window $Q_\delta\le Q\le Q_{\max}$, which is non-empty iff $Q_\delta\le Q_{\max}$. 

Within $[Q_\delta,Q_{\max}]$, a larger $Q$ means each good node holds more pieces, so a client needs to query fewer good nodes, $\hQ=\lceil\kappa/Q\rceil$, thus fewer round-trips (smaller $\Lambda^{\mathrm{succ}}_{\mathrm{get}}$). So is the decoding failure probability $P_2$. $P_1$ does not depend on $Q$. The latency $\Lambda_{\mathrm{get}}$ (Eq \eqref{eq:objective}) is smaller as $Q$ increases. This makes $Q =  Q_{\max}$ the optimal choice. 
\end{proof}

Based on the above corollary, different object types on the same grid may choose different $Q$ in the window independently, since the network is untouched. If monitoring shows $\Np$ falling, nodes raise $Q$ in place (Section~\ref{sec:tune}).
The next corollaries  are scaling laws assuming typical columns, with column size $N_c=N/K_2$. 

\begin{corollary}[Population limit]
\label{cor:carry}
On a grid with $K_2$ columns of $\Nbar=N/K_2$ nodes, TCDA at retention depth $Q$ meets the PUT budget iff network size $N\le(\kappa/Q)\,K_2^2B_{\mathrm P}/\sigma$.
\end{corollary}
\begin{proof}
On a typical column $c$, $N_c=\Nbar=N/K_2$, and so, applying Eq \ref{eq:PhiPut}, a relay in this column incurs a PUT cost of
\[
\Phi_{\mathrm{put}} = \frac{Q\sigma}{\kappa  K_2} N_c = \frac{Q}{\kappa}\,\frac{N\sigma}{K_2^2}.
\]
The PUT budget constraint $\Phi_{\mathrm{put}}\le B_{\mathrm P}$ is thus equivalent to $N\le(\kappa/Q)K_2^2B_{\mathrm P}/\sigma$.
\end{proof}

\begin{corollary}[Good-node range and object size limit]
\label{cor:region}
Let $N=\rho\Np$, $\delta_{\mathrm c} := 1-\sqrt{1-\delta}$, and  $x_1$ as in Theorem~\ref{thm:bootstrap} for $Q=1$. TCDA can be both $\delta$-robust and within the PUT budget with $\Np$ good nodes
\begin{enumerate}[leftmargin=*,label=(\roman*)]
\item only if (necessity)
\[
(K_1K_2-1)\ln(1/\delta_{\mathrm c})\ \le\ \Np\ \le\ \kappa K_2^2B_{\mathrm P}/(\rho\sigma).
\] 
\item if (sufficiency)
\[
\max\{K_1K_2\ln(4/\delta),\,K_2x_1\}\ \le\ \Np\ \le\ \kappa K_2^2B_{\mathrm P}/(\rho\sigma).
\]
\end{enumerate}
In particular, a feasible design exists only if 
\[
\sigma\ \le\ \sigma_{\max}:=\frac{\kappa K_2^2B_{\mathrm P}}{\rho\,(K_1K_2-1)\ln(1/\delta_{\mathrm c})}.
\]
\end{corollary}
\begin{proof}
Replacing $N=\rho\Np$ in Corollary~\ref{cor:carry}, the PUT budget requirement becomes 
\[
\Phi_{\mathrm{put}}(Q)=\frac{Q}{\kappa}\,\frac{\rho\Np\sigma}{K_2^2}\ \le\ B_{\mathrm P}\iff \Np\le\frac{\kappa}{Q}\,\frac{K_2^2B_{\mathrm P}}{r\sigma}.
\]

\emph{Necessity (i)} The above inequality immediately implies the upper bound $\Np\le\kappa K_2^2B_{\mathrm P}/(\rho\sigma)$ because $Q \ge 1$. The lower bound is resulted from the $\delta$-robustness, as follows. Because robustness is achieved by some $Q \le \kappa$, it must also be achieved if $Q=\kappa$ (because more pieces are stored, hence improving robustness). Applying this case to Theorem \ref{thm:bootstrap}, we have $\Np\ge\Np_\delta(\kappa) \ge (K_1K_2-1)\ln(1/\delta_{\mathrm c})$. The latter inequality is from Corollary~\ref{cor:qdelta}(ii).

\emph{Sufficiency (ii)} Take $Q=1$. The upper bound puts it within the PUT budget, as above. For robustness, $\hQ=\kappa\ge2$ at $Q=1$, so Eq.~\eqref{eq:NpdeltaQ} gives $\Np_\delta(1)\le\max\{K_1K_2\ln(4/\delta),K_2x_1\}\le\Np$. Applying $Q=1$ in Theorem~\ref{thm:bootstrap},  TCDA is $\delta$-robust. Thus $Q=1$ is feasible.

\emph{Bound on object size $\sigma$.} The range in (i) is non-empty iff $(K_1K_2-1)\ln(1/\delta_{\mathrm c})\le\kappa K_2^2B_{\mathrm P}/(\rho\sigma)$. Solving for $\sigma$ gives the bound. This bound caps the object size for every $\Np$.

\end{proof}

\textbf{Example.} With $\delta=1/4$, $\kappa=64$, $\rho := N/\Np = 2$, and $B_{\mathrm P}=9.3$\,MiB, the $(15,32)$ grid has $\sigma_{\max}\approx318$\,MiB for TCDA but only $\sigma_{\max} = 5$\,MiB for full replication (RDA). Ethereum's $32$\,MiB object therefore fits TCDA, for any $\Np \in [965, 9566]$, but not RDA on this grid at any network size.

\subsection{ Solution Algorithm} 
The steps of our design are sketched in Algorithm~\ref{alg:design}. For each $K_2$, the number of rows ranges over $K_{1,\min}\le K_1\le K_{1,\max}$, where $K_{1,\min}:=\lceil 1/(d_{\max}-1/K_2)\rceil$ is the degree constraint and $K_{1,\max}:=\lfloor(\Np/\ln(1/\delta_{\mathrm c})+1)/K_2\rfloor$ follows from robustness, $\Np\ge\Np_\delta(\kappa)\ge(K_1K_2-1)\ln(1/\delta_{\mathrm c})$ (Corollary~\ref{cor:qdelta}(ii)). On each grid, Corollary~\ref{cor:window} gives the optimal $Q$ in closed form,
so problem~\eqref{eq:opt} is solved by a finite scan over $(K_1,K_2)$. A scan is needed because $\Lambda_{\mathrm{get}}$ need not be monotone in $K_1$ or $K_2$.
On each grid, the robustness test is a single evaluation of $p_{\mathrm{fail}}$ at $Q_{\max}$, which suffices because $p_{\mathrm{fail}}$ is non-increasing in $Q$ (Corollary~\ref{cor:window}), so $Q_\delta\le Q_{\max}$ iff $p_{\mathrm{fail}}(Q_{\max})\le\delta$. The objective $\Lambda_{\mathrm{get}}$ is evaluated by simulating the GET, about one second per grid, or in closed form by the formula of Section~\ref{sec:getcost} with $\Lambda^{\mathrm{succ}}_{\mathrm{get}}$ at its bound.

\textbf{Complexity.} We count arithmetic operations, taking a power and a binomial CDF value as $O(1)$ each. The loop in line~\ref{ln:K2} runs over the divisors of $M$: at most $2\sqrt M$ of them, found in $O(\sqrt M)$ time. Per $K_2$, line~\ref{ln:Nc} takes $O(\log N_{\max})$ CDF evaluations by bisection, and lines~\ref{ln:QP}--\ref{ln:Qchk} take $O(1)$. The loop in line~\ref{ln:K1} runs at most $K_{1,\max}$ times. Per $(K_1,K_2)$, line~\ref{ln:pfail} costs $O(\hQ)=O(\kappa)$, the sum in $P_2$, and lines~\ref{ln:L}--\ref{ln:upd} cost $O(1)$. Since $\sum_{K_2\mid M}1/K_2=O(\log\log M)$, Algorithm~\ref{alg:design} runs in
\[
O\Big(\sqrt M\log N_{\max}+\kappa\big(\Np/\ln(1/\delta)+1\big)\log\log M\Big)
\]
operations,  $\approx 15,000$ for $M=128$, $\kappa=64$, and $\Np=5000$.

\begin{algorithm}[t]
\caption{Joint design of grid and retention depth}
\label{alg:design}
\begin{algorithmic}[1]
\Require
\Statex \textit{Object:} size $\sigma$, width $M$, layers $\kappa$
\Statex \textit{Network:} good nodes $\Np$, all nodes $N_{\max}$, degree $d_{\max}$
\Statex \textit{Budgets:} PUT $B_{\mathrm P}$ (quantile $\epsilon_{\mathrm P}$), storage $B_{\mathrm S}$, retention $W$
\Statex \textit{Robustness:} target $\delta$
\Ensure an optimal $(K_1,K_2,Q)$ of~\eqref{eq:opt}, or \textsc{infeasible}
\State $\mathcal D\gets\textsc{infeasible}$;\ \ $\Lambda^\star\gets\infty$
\For{each divisor $K_2$ of $M$ with $1/K_2<d_{\max}$}\label{ln:K2}
  \State $\hat N_c\gets(1-\epsilon_{\mathrm P})$-quantile of $\mathrm{Bin}(N_{\max},1/K_2)$\label{ln:Nc}
  \State\label{ln:QP} $Q_{\mathrm P}\gets\lfloor\kappa B_{\mathrm P}K_2/(\hat N_c\sigma)\rfloor$;\ \ $Q_{\mathrm S}\gets\lfloor\kappa B_{\mathrm S}K_2/(W\sigma)\rfloor$
  \State\label{ln:Q} $Q\gets\min\{\kappa,Q_{\mathrm P},Q_{\mathrm S}\}$ \Comment{Cor.~\ref{cor:window}}
  \If{$Q<1$}\label{ln:Qchk} \textbf{continue} \EndIf
  \For{$K_1=K_{1,\min},\dots,K_{1,\max}$}\label{ln:K1}
    \If{$p_{\mathrm{fail}}(K_1,K_2,Q)>\delta$}\label{ln:pfail} \textbf{continue} \Comment{Cor.~\ref{cor:sample}}
    \EndIf
    \State\label{ln:L} $\Lambda\gets\Lambda_{\mathrm{get}}(K_1,K_2,Q)$ \Comment{Sec.~\ref{sec:getcost}}
    \If{$\Lambda<\Lambda^\star$}
      \State\label{ln:upd} $\mathcal D\gets(K_1,K_2,Q)$;\ \ $\Lambda^\star\gets\Lambda$
    \EndIf
  \EndFor
\EndFor
\State \Return $\mathcal D$
\end{algorithmic}
\end{algorithm}

\section{Numerical Evaluation}
\label{sec:eval}

We evaluate TCDA at Ethereum's parameters for two purposes:  validate its theoretical results and  compare it to RDA and PeerDAS, the DAS design deployed on Ethereum. PeerDAS also represents the gossip-based approach, as opposed to the grid approach of TCDA and RDA. As in RDA, we report the GET cost at protocol level, in round-trips and bytes, and use wall-clock time only for the PUT, which has a deadline. The evaluation's source codes are available at \url{https://github.com/UMB-NCLab/tcda-sim}.

\textbf{Adversary.} Nodes, including publishers and  clients, are placed uniformly at random on the grid, and so are the sampled positions. A bad node is a Sybil, a dishonest node, or an offline honest node, and behaves in the worst case: a bad relay drops the chunk or forwards pieces with other coefficients, which the commitment check rejects; a bad collector returns nothing; and a queried bad node stays silent. In some scenarios, a mass of the good nodes also goes offline at once.

\subsection{Theoretical results at Ethereum's parameters}
\label{sec:eval-valid}
\label{sec:eval-setup}
We validate the results that rest on assumptions or give bounds; the others, such as Theorem~\ref{thm:bootstrap} and the corollaries of Section~\ref{sec:design}, are exact. The objectives are:
\begin{enumerate}[leftmargin=*,label=(O\arabic*)]
\item \emph{Growth of the GET cost with bad nodes}. PUT and storage stay within budget by design, so only the GET cost grows with bad nodes. We measure how, against the theoretical derivations.
\item \emph{Feasibility and latency of the optimal design}. Does Algorithm~\ref{alg:design} find a feasible design, how fast is its GET, and does the closed-form bound pick the same design as simulation?
\item \emph{Choosing the design inputs}. How do $\Np$, $N_{\max}$, and $d_{\max}$ change the design?
\end{enumerate}

\textbf{Settings.} Table~\ref{tab:params} lists the parameters and their justification. They follow Ethereum and the baselines' own evaluations, e.g., $M=128$ as in PeerDAS and $d_{\max}=1/10$ as in RDA. The storage budget, $32$\,GiB per node, is half of PeerDAS's ($0.5$\,MiB per object); it caps $Q$ at $K_2/2$.

\begin{table}[t]
\centering
\caption{Evaluation parameters.}
\label{tab:params}
\scriptsize
\setlength{\tabcolsep}{3pt}
\begin{tabular}{>{\raggedright\arraybackslash}p{1.95cm}>{\raggedright\arraybackslash}p{2.55cm}>{\raggedright\arraybackslash}p{3.7cm}}
\toprule
Parameter & Value & Justification\\
\midrule[0.8pt]
\multicolumn{3}{@{}l}{\textit{Object and coding}}\\
Field element; proof & $32$\,B; $\sigma_{\mathrm{pf}}=48$\,B & BLS12-381, as in EIP-4844~\cite{eip4844}\\
Symbol, $\kappa$ & $64$ elements ($2$\,KiB) & a PeerDAS cell~\cite{fulu-specs}\\
Object width $M$ & $128$ & PeerDAS columns~\cite{fulu-specs}\\
Object size $\sigma$ & $32$\,MiB & $64$ blobs of $128$\,KiB, 2D-extended\\
\midrule[0.8pt]
\multicolumn{3}{@{}l}{\textit{Robustness}}\\
Target $\delta$ & $1/4$ & $M=2D$ (Section~\ref{sec:goal})\\
\midrule[0.8pt]
\multicolumn{3}{@{}l}{\textit{Network}}\\
Population $N$, $N_{\max}$ & $2\Np$ (also $\Np$ to $16\Np$) & half bad; no effect on robustness (Thm.~\ref{thm:bootstrap})\\
Degree limit $d_{\max}$ & $1/10$ (also $1/6$) & RDA's example~\cite{rda}\\
\midrule[0.8pt]
\multicolumn{3}{@{}l}{\textit{Timing and budgets}}\\
Slot; attestation deadline & $12$\,s; $4$\,s & Ethereum~\cite{paradigm-slot}\\
Relay uplink & $25$\,Mbit/s (also $50$, $100$) & attesting node~\cite{eip7870}\\
PUT budget $B_{\mathrm P}$ & $9.3$\,MiB & $4$\,s at $25$\,Mbit/s minus the publisher's $0.87$\,s\\
PUT sizing $\epsilon_{\mathrm P}$ & $0.01$ & $99$th-percentile column (Section~\ref{sec:designproblem})\\
Retention $W$; storage $B_{\mathrm S}$ & $131{,}072$ blocks; $32$\,GiB & $18$ days~\cite{eip4844}; half of PeerDAS's per-node storage\\
\bottomrule
\end{tabular}
\end{table}

\subsubsection{Objective O1 - Growth of the GET cost with bad nodes}
\label{sec:eval-cost}
\begin{figure}[t]
\centering
\includegraphics[width=222.52bp]{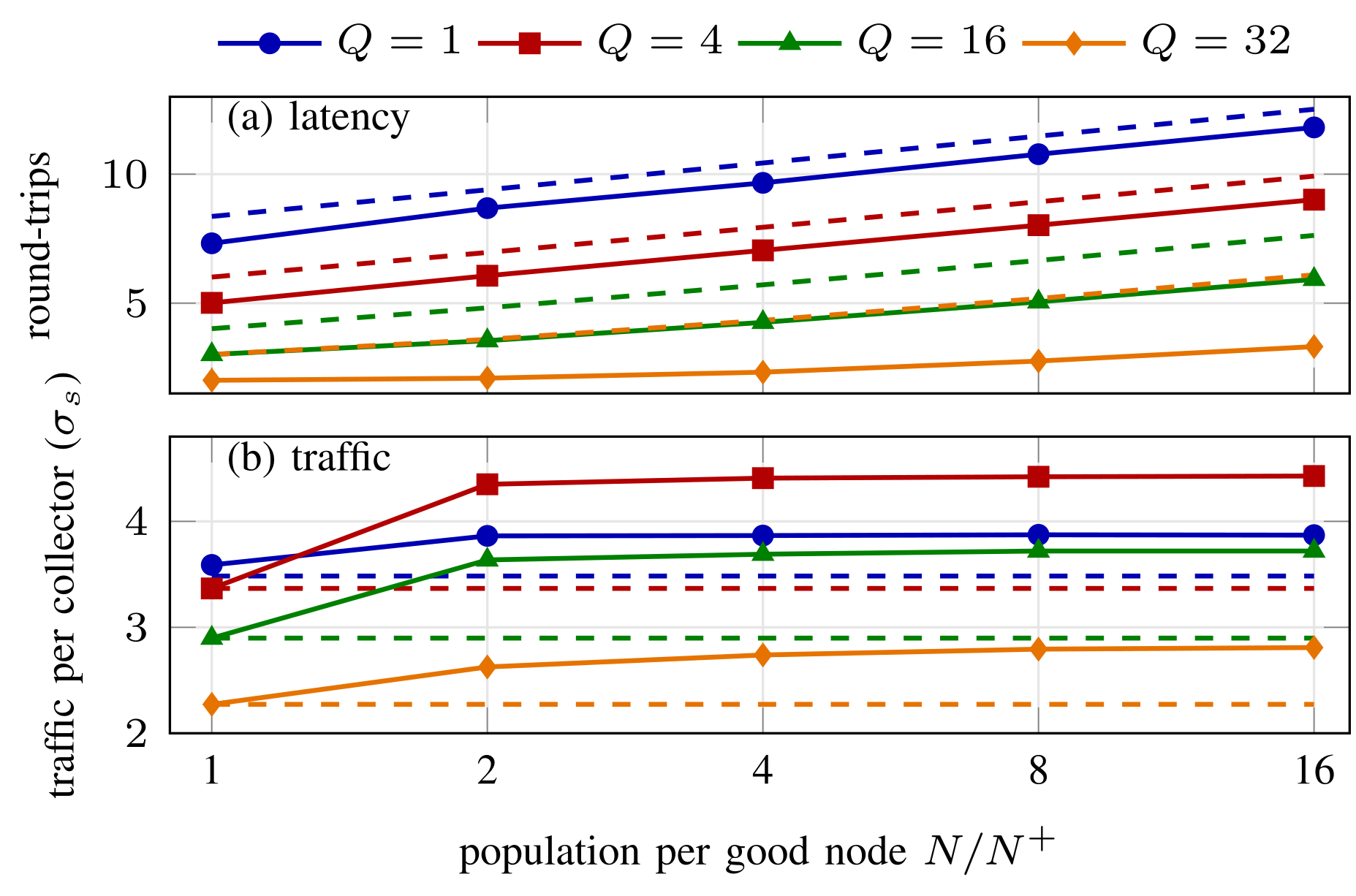}
\caption{(O1) GET cost per sample, failed attempts included, as bad nodes join, measured (solid) and the bounds of Section~\ref{sec:getcost} (dashed): (a) round-trips, with $\Lambda^{\mathrm{succ}}_{\mathrm{get}}$ at its bound; (b) traffic sent by good nodes per collector, against its minimum $(\kappa-Q)(1/\kappa+\beta)+1+\beta$. $(15,32)$ grid, $\Np=2500$.}
\label{fig:growth}
\end{figure}

Figure~\ref{fig:growth} shows the GET cost as bad nodes join: $4000$ samples per point, failed attempts included, on the $(15,32)$ grid with $\Np=2500$, which is $\delta$-robust for every tested $Q$, and $0$ to $15$ bad nodes per good one. The latency grows logarithmically, as derived (Figure~\ref{fig:growth}a). At $Q=4$, each doubling of $N/\Np$ adds one round-trip ($5.0$, $6.1$, $7.0$, $8.0$, and $9.0$ at $N/\Np=1$ to $16$), the slope of the formula, which stays about one round-trip above. At large $Q$ the growth is slower than the formula, which is per collector, while a client takes the fastest of its cell's collectors, about five here. The traffic per collector does not grow (Figure~\ref{fig:growth}b): it stays within $31\%$ of its minimum, since silent bad peers send nothing.

\subsubsection{Objective O2 - Feasibility and latency of the optimal design}
\label{sec:eval-design}
We apply Algorithm~\ref{alg:design} at uplinks of $25$, $50$, and $100$\,Mbit/s, $\Np$ up to $40{,}000$, and $N_{\max}=2\Np$. The PUT and storage costs are exact, so only $\Lambda_{\mathrm{get}}$ is simulated, with $1500$ samples, about one second per grid. The closed-form bound picks the same design as simulation in all $69$ cases, so a deployment can choose its design without simulation.

\emph{Feasibility.} A feasible design exists for every $\Np\ge868$ at all three uplinks, and for every $\Np\ge330$ at $d_{\max}=1/6$ (Table~\ref{tab:design}). Storage binds only at the faster uplinks.

\begin{figure}[t]
\centering
\includegraphics[width=225.416bp]{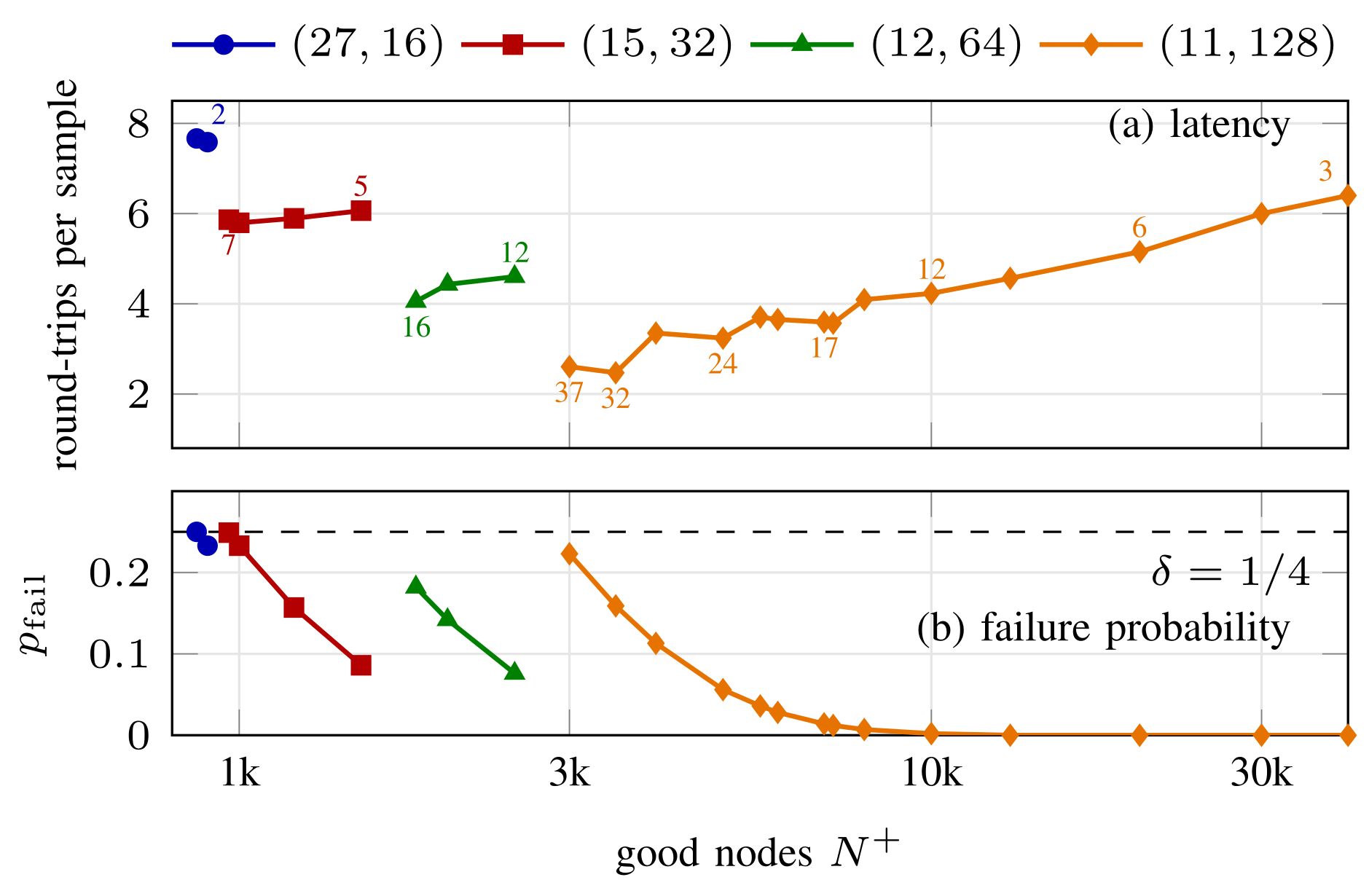}
\caption{(O2, O3) The design chosen by Algorithm~\ref{alg:design} at $25$\,Mbit/s, colored by its grid $(K_1,K_2)$ ($d_{\max}=1/10$, $N_{\max}=2\Np$, $\epsilon_{\mathrm P}=0.01$, $\sigma=32$\,MiB, $\delta=1/4$): (a) round-trips per sample, with $Q$ written next to selected points; (b) failure probability per sample. No design fits below $868$ good nodes.}
\label{fig:design}
\end{figure}

\emph{Latency.} Figure~\ref{fig:design}a shows the latency of the chosen design at $25$\,Mbit/s. As predicted, once $\Np$ grows enough, the design moves to a grid with twice as many columns, where the PUT budget allows a larger $Q$ and a faster GET: the latency falls from $7.7$ to $2.6$ round-trips up to about $3000$ good nodes. Beyond that, $K_2=M=128$, so the columns fill and the latency grows logarithmically.

\begin{table}[t]
\centering
\caption{(O2, O3) Good-node ranges with a feasible design, and its round-trips per sample $\Lambda_{\mathrm{get}}$ (Algorithm~\ref{alg:design}, $N_{\max}=2\Np$, $\epsilon_{\mathrm P}=0.01$, $\sigma=32$\,MiB, $\delta=1/4$, $\Np\le40{,}000$).}
\label{tab:design}
\scriptsize
\setlength{\tabcolsep}{3pt}
\begin{tabular}{@{}rcccc@{}}
\toprule
 & \multicolumn{2}{c}{$d_{\max}=1/10$} & \multicolumn{2}{c}{$d_{\max}=1/6$}\\
\cmidrule(lr){2-3}\cmidrule(lr){4-5}
Uplink & Feasible $\Np$ & $\Lambda_{\mathrm{get}}$ & Feasible $\Np$ & $\Lambda_{\mathrm{get}}$\\
\midrule
25\,Mbit/s & ${\ge}868$ & 2.5--7.7 & ${\ge}330$ & 2.2--7.3\\
50\,Mbit/s & ${\ge}868$ & 1.2--6.6 & ${\ge}321$ & 1.0--5.6\\
100\,Mbit/s & ${\ge}868$ & 1.0--5.5 & ${\ge}321$ & 1.0--5.5\\
\bottomrule
\end{tabular}
\end{table}

\begin{figure}[t]
\centering
\includegraphics[width=246.919bp]{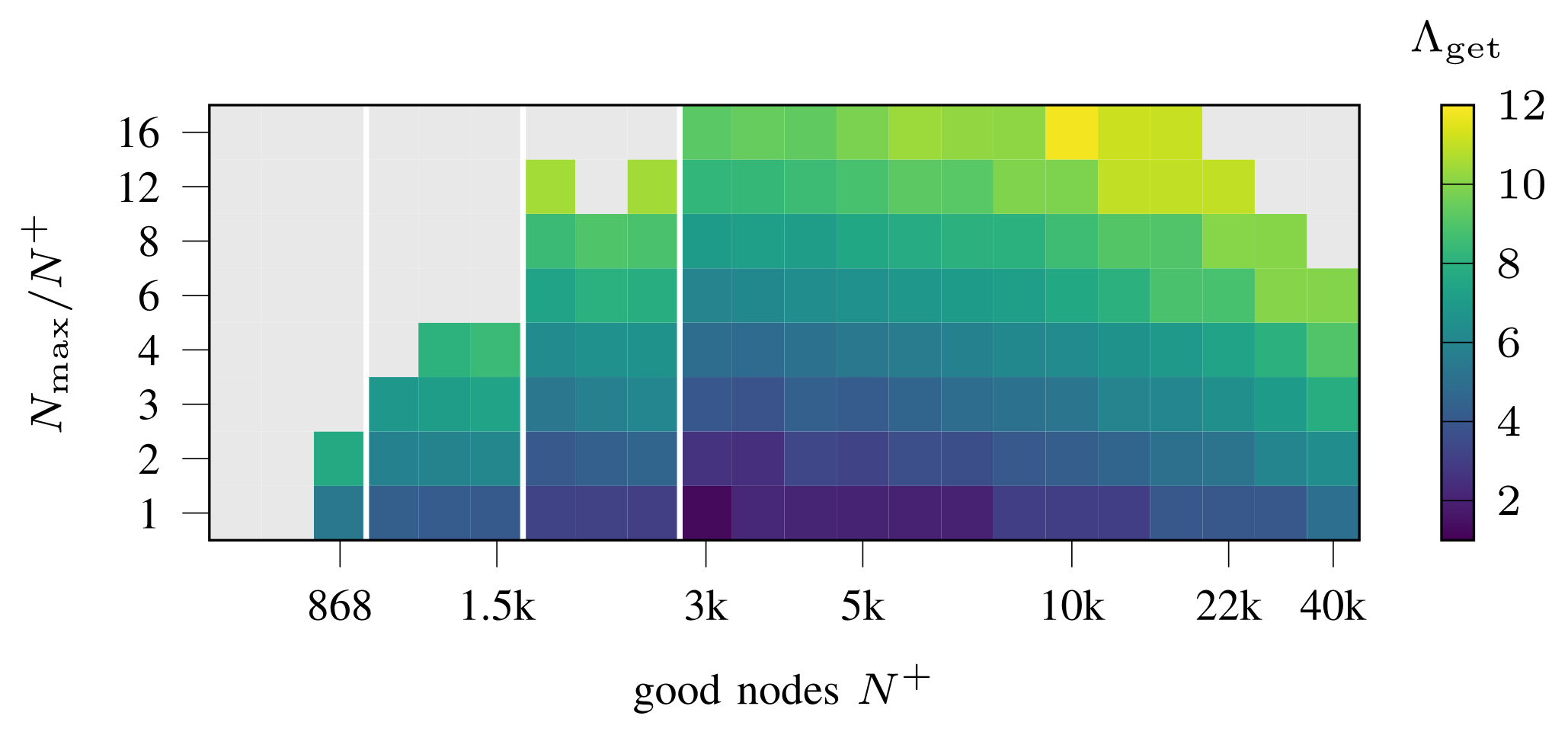}
\caption{(O3) GET latency $\Lambda_{\mathrm{get}}$ of the design chosen by Algorithm~\ref{alg:design} over both inputs, the good nodes $\Np$ and the ratio $N_{\max}/\Np$, at $25$\,Mbit/s ($d_{\max}=1/10$). Grey: no feasible design. White lines: the chosen grid changes, to $32$, $64$, and $128$ columns.}
\label{fig:ratio}
\end{figure}

\subsubsection{Objective O3 - Choosing the design inputs}
\label{sec:eval-dmax}
Algorithm~\ref{alg:design} takes $\Np$, $N_{\max}$, and $d_{\max}$ as inputs. A deployment prefers a small $\Np$, which is easier to reach, a $p_{\mathrm{fail}}$ safely below $\delta$ to tolerate churn, and a fast GET.

\emph{Bootstrap size.} With $N_{\max}/\Np=2$, the fastest size, about $3000$ to $3500$ good nodes, has a thin margin: $p_{\mathrm{fail}}$ is $0.16$ to $0.22$ against $\delta=0.25$ (Figure~\ref{fig:design}b). A little above, the margin grows quickly at a small cost: at $5000$ good nodes, $p_{\mathrm{fail}}=0.056$, four times below $\delta$, for $3.2$ round-trips instead of $2.5$.

\emph{Population bound.} Figure~\ref{fig:ratio} varies $N_{\max}/\Np$. The pattern of Figure~\ref{fig:design}a holds at every ratio, and a larger ratio costs twice. The GET is slower, by about two round-trips per doubling: $2.6$, $4.9$, $7.1$, and $9.2$ at $3000$ good nodes and ratios $2$, $4$, $8$, and $16$. The smallest feasible network is larger: $868$, $1200$, $1800$, and $3000$ good nodes. No design exists beyond about $300{,}000$ nodes in total, where even $Q=1$ exceeds the PUT budget. The fastest designs lie at about $3000$ to $7000$ good nodes and a ratio of at most $2$: the best $\Np$ is at neither end of its range, while a smaller ratio is faster but leaves less room for growth and bad nodes.

\emph{Degree limit.} With $d_{\max}=1/6$, the bootstrap drops from $868$ to about $330$ good nodes, since fewer, larger cells rarely lack a good node. The GET is faster by up to four round-trips below $3000$ good nodes and by at most $0.3$ above. The degree limit trades a node's connections against the bootstrap, not against feasibility.

\subsection{Comparison with the baselines}
\label{sec:eval-bench}
Table~\ref{tab:headline} summarizes the comparison at $\Np=2000$ in the default setting. PeerDAS has the lightest per-node load but no robustness guarantee, and fails against Sybils. RDA has the fastest GET but misses the PUT deadline. TCDA is $\delta$-robust once it has enough good nodes, whatever the number of bad ones, and within budget while the network stays within $N_{\max}$. The objectives are:
\begin{enumerate}[leftmargin=*,label=(O\arabic*),resume]
\item \emph{Feasibility and robustness}. At which network sizes does each design meet the budgets, and does it stay $\delta$-robust under Sybils?
\item \emph{PUT time with coding included}. With the publisher's row hop and the relays' coding time, is the deadline still met?
\item \emph{Dynamic network}. Our analysis is static. Is $\delta$-robustness kept as nodes join and leave, or as many leave at once?
\end{enumerate}

\begin{table*}[t]
\centering
\caption{(O4, O5) Overall comparison of the three designs at $\Np=2000$ good nodes and as many bad ones, $\sigma=32$\,MiB ($8$\,MiB of blob data), $\delta=1/4$, $25$\,Mbit/s uplink, $4$\,s deadline, $d_{\max}=1/10$. PUT sized on the $99$th-percentile column; PUT in time from Section~\ref{sec:eval-put}; GET round-trips per sample, failed attempts included.}
\label{tab:headline}
\footnotesize
\setlength{\tabcolsep}{3.5pt}
\begin{tabular}{@{}llcclccc@{}}
\toprule
Design & Robust iff & $p_{\mathrm{fail}}$, $0\to10\Np$ Sybils & Min.\ $\Np$ for 4\,s & Grid, $Q$ & PUT in time & Storage/node & GET round-trips\\
\midrule
PeerDAS & no guarantee & $0\to0.91$ & load indep.\ of $N$ & 128 columns & --- & $0.5$\,MiB & $1$ (req./resp.)\\
RDA & $\Np\ge\Np_\delta(\kappa)$ & $0.14\to0.14$ & none & $(12,64)$, $64$ & $0\%$ & $0.5$\,MiB & $1.2$\\
\textbf{TCDA} & $\Np\ge\Np_\delta(Q)$ & $0.14\to0.14$ & $\mathbf{868}$ & $(12,64)$, $14$ & $99.6\%$ & $112$\,KiB & $4.4$\\
\bottomrule
\end{tabular}
\end{table*}

\textbf{Settings.} The parameters are those of Table~\ref{tab:params}. \emph{PeerDAS}~\cite{peerdas,fulu-specs,fusaka}: blobs are 1D-erasure-coded into $M=128$ columns, its chunks. Each node stores at least $4$ and samples at least $8$ by subscribing to their gossip subnets. Custody is self-advertised, so any node can claim any chunk. \emph{RDA}~\cite{rda} is TCDA at $Q=\kappa$, except for the condition on the good nodes. We count an RDA grid as feasible if $\Np\ge\Np_\delta(\kappa)$, which is less demanding than RDA's own per-column assumption at $\epsilon=10^{-9}$ and so favors RDA.

\subsubsection{Objective O4 - Feasibility and robustness}
\label{sec:eval-sota}

\begin{figure}[t]
\centering
\includegraphics[width=230.548bp]{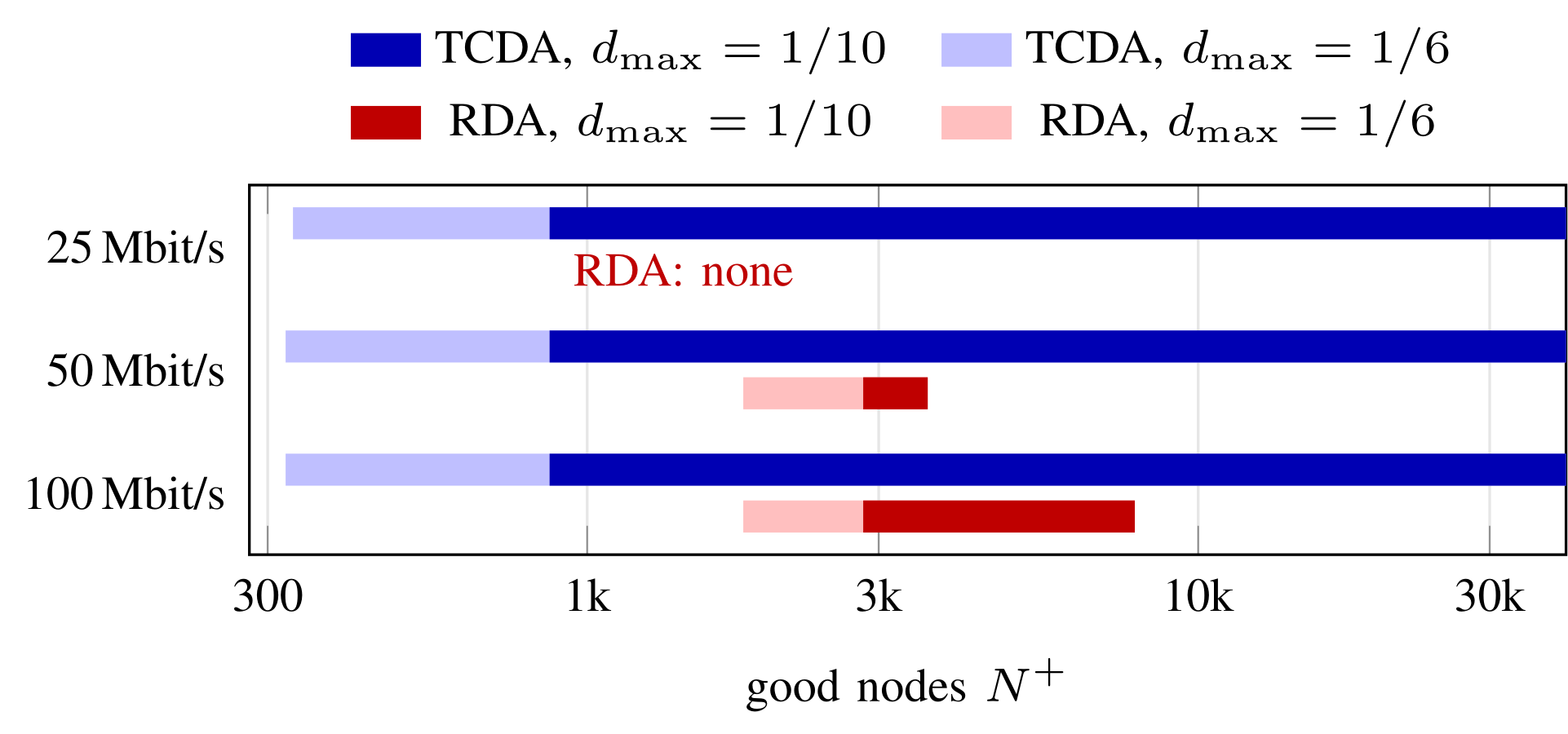}
\caption{(O4) Network sizes at which a design meets the PUT and storage budgets and is $\delta$-robust ($N_{\max}=2\Np$, $\sigma=32$\,MiB, $\delta=1/4$, $\Np\le40{,}000$). The bars for $d_{\max}=1/6$ show only the sizes it adds to those of $d_{\max}=1/10$.}
\label{fig:feasible}
\end{figure}

Figure~\ref{fig:feasible} shows the network sizes at which each design meets the budgets. TCDA meets them at every size from its bootstrap on, at every uplink. At $25$\,Mbit/s, RDA fails on every grid; a looser degree limit lowers its bootstrap but not its load. At faster uplinks, RDA is feasible only on $128$ columns, where a full chunk fits the storage budget, and only until its load, which grows with the network, exceeds the PUT budget. PeerDAS meets the budgets at every size, as its per-node load does not depend on the network, but it fails robustness.

\begin{figure}[t]
\centering
\includegraphics[width=249.248bp]{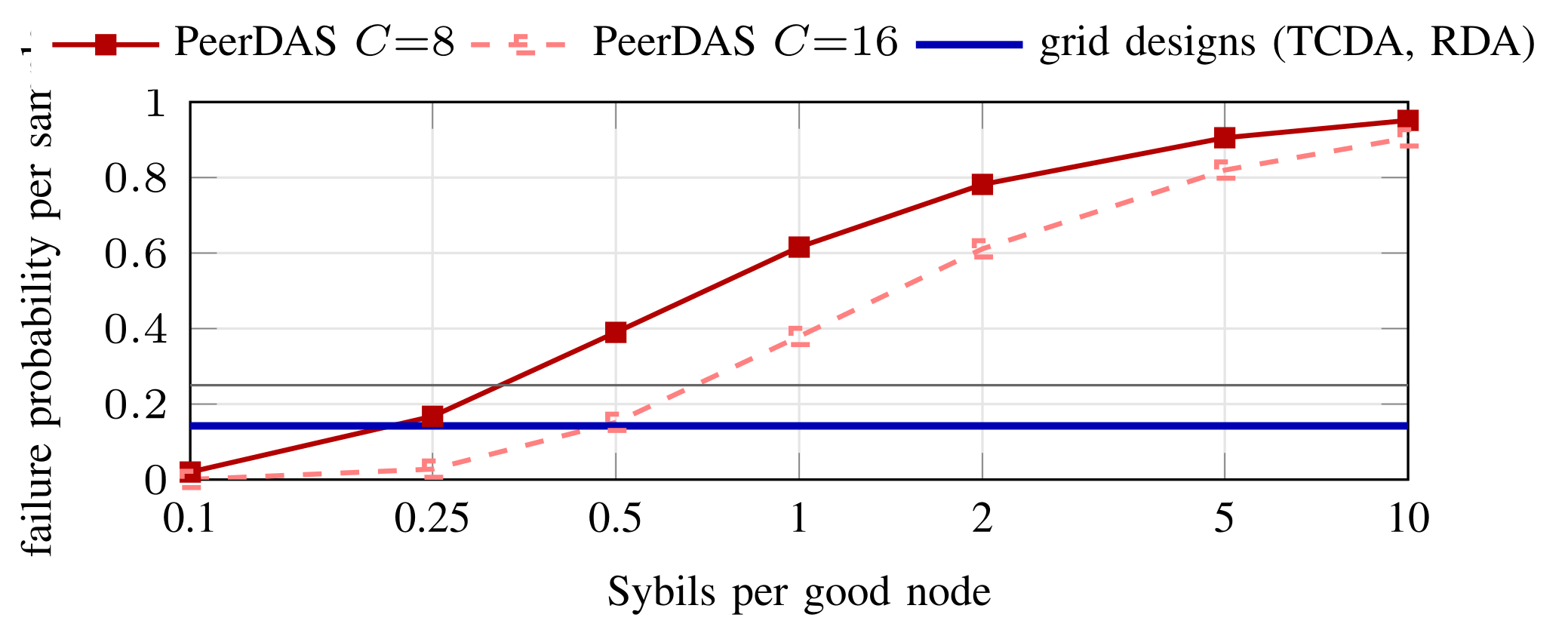}
\caption{(O4) Per-sample failure probability at $\Np=2000$ as Sybils join. PeerDAS: Sybils advertise custody of every chunk; a node contacts $C$ advertisers. The grid designs (RDA and TCDA on $(12,64)$, Table~\ref{tab:headline}) are unaffected: their failure probability depends on the good nodes only. Thin line: $\delta=1/4$.}
\label{fig:sybil}
\end{figure}

\textbf{Sybil attack.}\label{sec:eval-sybil} In PeerDAS, each node advertises the chunks it stores, and a node that samples a chunk receives it from its $8$ gossip peers for that chunk. We let the Sybils advertise every chunk and, optimistically for PeerDAS, let every good custodian hold its chunks. A sample contacts $C\in\{8,16\}$ distinct advertisers, drawn uniformly, and fails if all of them are Sybils. Figure~\ref{fig:sybil} shows the result at $\Np=2000$. PeerDAS loses robustness from $0.33$ Sybils per good node with $C=8$ and $0.7$ with $C=16$, and worsens as more join, while TCDA and RDA keep their exact $p_{\mathrm{fail}}$ at every Sybil count. Peer rating and IP limits raise the cost of a Sybil attack but do not bound its effect, so PeerDAS is not fit where robustness comes first.
  
\textbf{Targeted attack.} Suppose the adversary can put bad nodes in a chosen column. We take the two grid designs of Table~\ref{tab:headline} and move all $2000$ bad nodes into one column. Robustness is unaffected: the failure rates are $0.145$ for TCDA and $0.144$ for RDA, against the exact $0.142$. RDA's GET stays at one hop, and TCDA's slows only logarithmically, from $4.3$ to $8.9$ round-trips per attempt. The attacker gains load: the relays of that column must serve all its $2031$ nodes, $32$ times the usual, and miss the deadline in both designs. This cannot be avoided, but breaking TCDA's deadline guarantee takes $\kappa/Q$ times more bad nodes in one column than breaking RDA's. In practice, cells should be assigned from randomness unknown when node IDs are created, to prevent the attack.

\subsubsection{Objective O5 - PUT time with coding included}
\label{sec:eval-put}
We now time the whole PUT, with the publisher's row hop and the relays' coding included. The publisher sends the chunks at $1$\,Gbit/s in total. Each good relay computes the coded pieces for each column node, at $64$\,ns per multiply-add in the scalar field of BLS12-381 (Montgomery arithmetic, measured on a virtual machine with $2$ cores of a $2.8$\,GHz Intel Xeon), and uploads them at $25$\,Mbit/s in random order. Coding runs ahead of the upload, since one core codes at about $60$\,Mbit/s. For each design and size, we time one random column in each of $300$ random networks. Table~\ref{tab:put} gives the 99th percentile of that time and the fraction of columns served within $4$\,s, on the same grid for both designs. TCDA meets the deadline in $98\%$ or more of the columns, and its 99th percentile, $3.8$ to $4.3$\,s, misses it by at most a few tenths of a second. RDA meets it in at most $2\%$: its relays need $9$ to $20$\,s. We do not time PeerDAS: it has no coding, and its load does not depend on the network size, so neither effect applies.

\begin{table}[t]
\centering
\caption{(O5) PUT time with coding included: $N=2\Np$, $\sigma=32$\,MiB, $25$\,Mbit/s, $d_{\max}=1/10$: $99$th percentile over columns of the time to serve a column (s), and the fraction of columns served within $4$\,s, relay coding included.}
\label{tab:put}
\scriptsize
\setlength{\tabcolsep}{3pt}
\begin{tabular}{@{}rcrrrrr@{}}
\toprule
 & & \multicolumn{3}{c}{TCDA} & \multicolumn{2}{c}{RDA}\\
\cmidrule(lr){3-5}\cmidrule(lr){6-7}
$\Np$ & Grid & $Q$ & p99 & met & p99 & met\\
\midrule
2000 & $(12,64)$ & 14 & 3.8 & 99.6\% & 12.8 & 0\%\\
6000 & $(11,128)$ & 20 & 4.2 & 98\% & 9.4 & 2\%\\
20{,}000 & $(11,128)$ & 6 & 4.3 & 99\% & 20.3 & 0\%\\
\bottomrule
\end{tabular}
\end{table}

\subsubsection{Objective O6 - Dynamic network}
\label{sec:eval-churn}
We consider steady churn and a sudden mass departure of good nodes. For churn, we follow RDA's simulation: the network grows from $20$ good nodes by one per step to $N^+_{\mathrm H}$; then, at every step, the $50$ oldest nodes leave and $50$ new ones join, so a node stays $N^+_{\mathrm H}/50$ steps, one \emph{lifetime}. At the end of this warm-up, we publish one object and follow its availability for $72$ lifetimes, Ethereum's retention window of $18$ days with a node online for $6$ hours~\cite{rda}. The grid is $(12,64)$, with $N^+_{\mathrm H}\in\{1600,2500,4000\}$, above its bootstrap of $1543$, and $5$ seeds. We test the optimal $Q_{\max}$ (Corollary~\ref{cor:window}) and the smallest $\delta$-robust $Q_\delta$ (Corollary~\ref{cor:qdelta}), which has the lowest PUT and storage costs but barely enough good nodes, so little margin against churn.

\emph{Sustainability of $\delta$-robustness.} TCDA at the optimal $Q$ behaves like RDA in every case (Table~\ref{tab:churn}). At $N^+_{\mathrm H}=2500$ and $4000$, both stay $\delta$-robust over the whole window, with worst failure rates of $0.12$ and $0.06$. At $1600$, just above the bootstrap, both are $\delta$-robust $60\%$ of the time, with a worst rate of $0.32$; in practice, both need some distance from the bootstrap. With $Q_\delta$ instead, at $1600$ (TCDA/$Q{=}4$) and $2500$ (TCDA/$Q{=}2$), the failure rate exceeds $\delta$ within $0.6$ lifetimes and every column loses the object; at $4000$, TCDA/$Q{=}2$ stays robust, with a worst rate of $0.06$. The optimal $Q$ thus gives a safe margin that $Q_\delta$ lacks.

\begin{figure}[t]
\centering
\includegraphics[width=233.156bp]{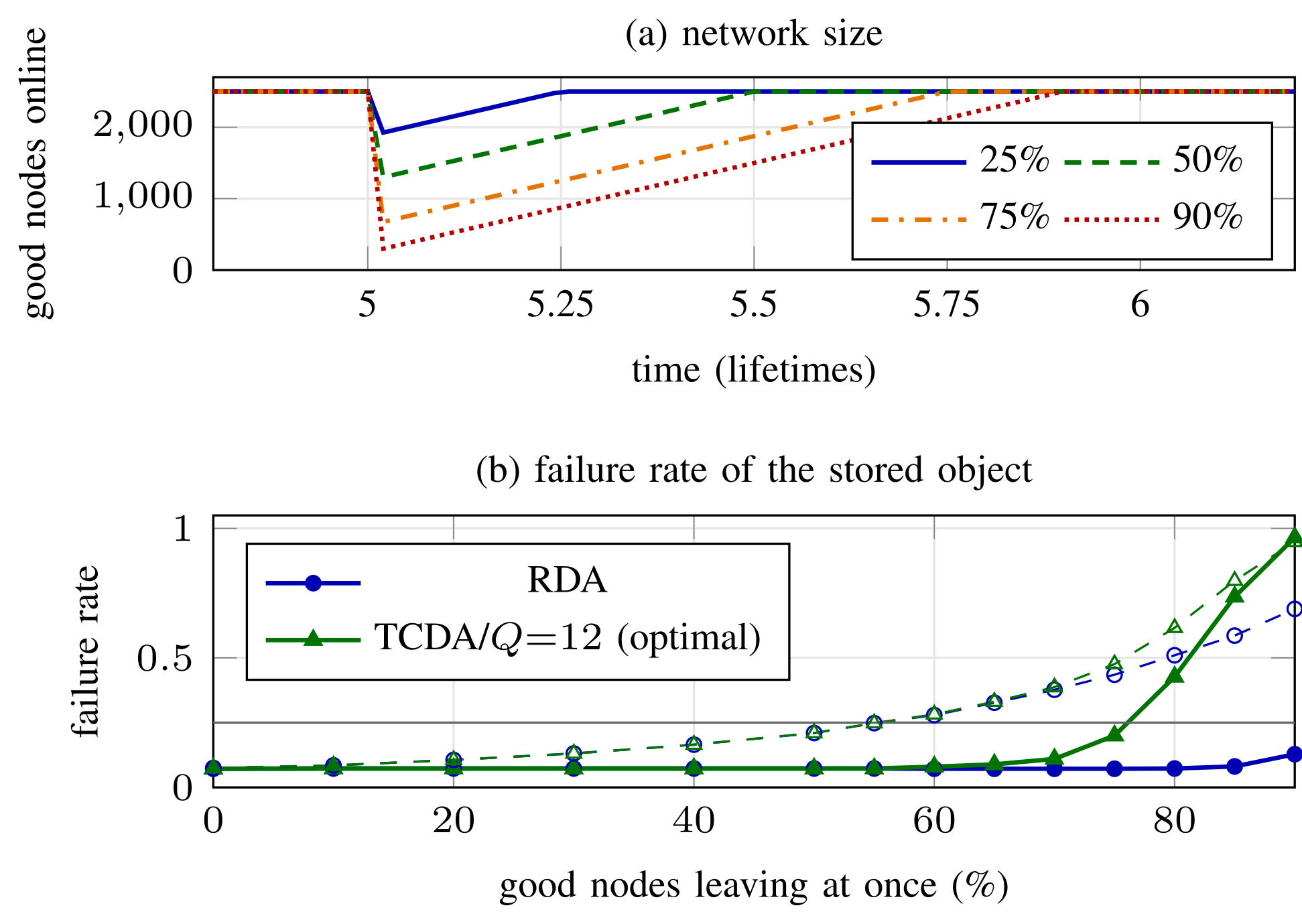}
\caption{(O6) An outage under churn, $(12,64)$ grid, $N^+_{\mathrm H}=2500$: (a) network size when $25$ to $90\%$ of the good nodes leave at once after $5$ lifetimes; (b) failure rate of the stored object, mean of $10$ seeds, after recovery (solid, filled marks) and right after the outage (dashed, open marks). Thin line: $\delta=1/4$.}
\label{fig:outage}
\end{figure}

\emph{Mass departures of good nodes.} After $5$ lifetimes, $25\%$ to $90\%$ of the good nodes, chosen at random, leave at once, and the network then refills by $50$ nodes per step (Figure~\ref{fig:outage}a). Figure~\ref{fig:outage}b shows the failure rate of the stored object at $N^+_{\mathrm H}=2500$, right after the outage and after recovery. Up to $55\%$, TCDA and RDA degrade but remain $\delta$-robust, and recover fully. At $75\%$, TCDA keeps a permanent loss but stays below $\delta$ after recovery, at $0.20$. Near $90\%$, it loses the object for good, while RDA recovers: one surviving good holder suffices for RDA, whereas TCDA needs several to decode.
\begin{table}[t]
\centering
\caption{(O6) A stored object under RDA's churn schedule, $(12,64)$ grid, $\delta=1/4$, $72$ lifetimes, $5$ seeds. ``Robust'': fraction of the window in which its failure rate is at most $\delta$; ``worst'': the largest failure rate; ``lost'': columns without $\hQ$ holders at the end, of $64$.}
\label{tab:churn}
\scriptsize
\setlength{\tabcolsep}{3pt}
\begin{tabular}{@{}rlrrrr@{}}
\toprule
$N^+_{\mathrm H}$ & Design & $\hQ$ & Robust & Worst & Lost\\
\midrule
1600 & RDA & 1 & $60\%$ & $0.32$ & 4--12\\
 & TCDA/$Q{=}17$ (optimal) & 4 & $60\%$ & $0.32$ & 4--12\\
 & TCDA/$Q{=}4$ ($Q_\delta$) & 16 & $0\%$ & $1.00$ & 64\\
\midrule
2500 & RDA & 1 & $100\%$ & $0.12$ & 1--4\\
 & TCDA/$Q{=}12$ (optimal) & 6 & $100\%$ & $0.12$ & 1--4\\
 & TCDA/$Q{=}2$ ($Q_\delta$) & 32 & $0\%$ & $1.00$ & 64\\
\midrule
4000 & RDA & 1 & $100\%$ & $0.06$ & 0--3\\
 & TCDA/$Q{=}7$ (optimal) & 10 & $100\%$ & $0.06$ & 0--3\\
 & TCDA/$Q{=}2$ ($Q_\delta$) & 32 & $100\%$ & $0.06$ & 0--3\\
\bottomrule
\end{tabular}
\end{table}

\section{Discussion}
\label{sec:discussion}

TCDA has limitations. Some cannot be avoided but are justified, and some can be addressed. We also see room for extensions beyond this work.

\subsection{Limitations}

A GET in TCDA takes a second hop into the column: a few round-trips, where RDA needs about one. This is the price of coding and cannot be avoided. We accept it because the PUT, not the GET, is on the critical path: a late PUT delays attestation and slows the whole blockchain, while a slow sample delays only the node that takes it. Also, after a mass departure of more than about $75\%$ of the good nodes, TCDA does not recover its robustness, whereas RDA does. We expect such an outage to be rare in practice. Finally, TCDA's design needs a bound on the largest network size as input. RDA is limited in the same way, but its bound is much tighter than ours.

\subsection{Open problems}

\textbf{A faster GET.} The second hop is what slows sampling, so shortening it is the first priority. A collector could query more nodes at once, trading traffic for round-trips, or remember which nodes answered earlier samples and ask those first. We have not studied these options.

\textbf{Repair in mass outage.} To address the outage limitation, we can extend TCDA such that good nodes in an affected column can rebuild pieces from the other columns, using the two-dimensional erasure code. This would give TCDA the same tolerance to severe outages as RDA.

\textbf{Beyond the worst case.} TCDA sends to every node of a column because it assumes nothing about where the good nodes are. A mild assumption on each column would allow sending to only part of it, and make the PUT cost independent of the network size.

\textbf{Network size as a design choice.} Our design takes the number of good nodes and the largest network size as given. Our evaluation suggests that both can be chosen as design variables to make sampling faster; choosing them jointly is an open problem.

\textbf{More routes into a column.} A GET fails if its contact cell has no good node. Through a bootstrap node, present in every row and good by assumption, the sampling node could instead use the contact cell of another row. This lowers the failure rate at no cost to the PUT, hence fewer good nodes needed to start.

\textbf{Realistic dynamics.} Our dynamic-network model can be made more realistic by studying real session times, targeted departures, and the wall-clock time of sampling. A real-world prototype would be ideal.


\section{Conclusion}
\label{sec:conclusion}
A trustless DAS layer must stay robust against any number of Byzantine peers and still deliver each block before validators attest. RDA, the only design with this guarantee, obtains it by full replication, but its PUT cost cannot be tuned, grows with the network, and misses Ethereum's attestation deadline at the block size we study. This paper shows that coding can replace replication. We have proposed TCDA, the first provably robust coded DAS layer, together with the exact condition under which it is robust. TCDA trades the PUT and storage costs against the GET cost on the same network. At Ethereum's parameters, this tuning covers a 64-fold range of costs, which serves a wide range of deployment constraints, and TCDA meets the attestation deadline where RDA does not. 
It maintains availability through an outage of up to about $55\%$ of the good nodes and regains robustness once the network refills after an outage of up to $75\%$. TCDA stays robust under any number of Sybils, where PeerDAS, in our model, does not. The price is a slower GET than RDA's. Speeding it up is the first item of our future work, followed by extensions that lower the bootstrap size and relax the bound on the network size, and by real-world experiments.


\bibliographystyle{IEEEtran}
\bibliography{tcda}

\end{document}